\documentclass[12pt]{article}
\usepackage{epsfig}
\usepackage{float}

\date{}

\newcommand{\ba}{\begin{array}}
\newcommand{\ea}{\end{array}}
\newcommand{\bd}{\begin{displaymath}}
\newcommand{\ed}{\end{displaymath}}
\newcommand{\bi}{\begin{itemize}}
\newcommand{\ei}{\end{itemize}}
\newcommand{\benu}{\begin{enumerate}}
\newcommand{\eenu}{\end{enumerate}}
\newcommand{\be}{\begin{equation}}
\newcommand{\ee}{\end{equation}}
\newcommand{\bea}{\begin{eqnarray}}
\newcommand{\eea}{\end{eqnarray}}
\usepackage{amsfonts}
\usepackage{multirow}
\usepackage{amssymb}
\usepackage[centertags]{amsmath}
\usepackage{graphicx}
\usepackage{hyperref}
\usepackage{booktabs}
\usepackage{theorem}
\usepackage{array}
\usepackage{epsfig}
\usepackage{colordvi}
\usepackage{color}
\usepackage{ulem}
\usepackage{cite}
\usepackage{color}
\usepackage{subfigure}
\usepackage{pdflscape}

\def\1{\mathbf{1}}
\def\3{\mathbf{3}}
\def\2{\mathbf{2}}

\newcommand{\bec}{\begin{cases}}
\newcommand{\eec}{\end{cases}}
\newcommand{\beq}{\begin{equation*}}
\newcommand{\eeq}{\end{equation*}}

\newcommand{\review}[1]{{\color{black}#1\color{black}}}

\makeatletter

\newcommand{\Rmnum}[1]{\expandafter\@slowromancap\romannumeral #1@}
\makeatother

\begin{document}

\begin{titlepage}
%

\begin{center}
{\bf{\Large Neutrino Oscillation Tomography of the Earth\\
\vspace*{0.2cm}
with the Hyper-Kamiokande Detector 
}}\\

\vspace{0.4cm} 
C. Jes\'us-Valls$\mbox{}^{a)}$\footnote{\review{cesar.jesus-valls@ipmu.jp\color{black}}} , 
S. T. Petcov$\mbox{}^{a,b)}$
\footnote{Also at: Institute of Nuclear Research and Nuclear Energy,
Bulgarian Academy of Sciences, 1784 Sofia, Bulgaria.} 
 and J. Xia$\mbox{}^{a)}$ 
\\[1mm]
\end{center}
\vspace*{0.50cm}
\centerline{$^{a}$ \it Kavli IPMU (WPI), UTIAS, The University of Tokyo,
Kashiwa,  Chiba 277-8583, Japan}
\vspace*{0.2cm}
\centerline{$^{b}$ \it INFN/SISSA, Via Bonomea 265, 34136 Trieste, Italy}
\vspace*{0.8cm}

\begin{abstract}
\noindent 

Using PREM as a reference model for the Earth density distribution we investigate the sensitivity of the Hyper-Kamiokande (HK) detector to deviations of the Earth i) core average density $\bar{\rho}_C$, ii) lower mantle average density $\bar{\rho}_{lman}$) and iii) upper mantle average density $\bar{\rho}_{uman}$, from their respective PREM densities. The analysis is performed by studying the effects of the Earth matter on the oscillations of atmospheric $\nu_{\mu}$, $\nu_e$, $\bar{\nu}_\mu$ and $\bar{\nu}_e$. We implement the constraints on the variations of $\rho_C$, $\rho_{lman}$ and $\rho_{uman}$ following from the precise knowledge of the Earth mass $M_\oplus$ and moment of inertia $I_\oplus$, as well as from the requirement that the Earth be in hydrostatic equilibrium (EHE). These constraints limit in the case of the three layer Earth density structure we are considering the maximal positive deviation of $\bar{\rho}_C$ from its PREM value to $10\%$. Considering the case of normal ordering (NO) of neutrino masses, we present results which illustrate the dependence of sensitivity to the core, lower and upper mantle average densities on the energy and zenith angle resolutions and on the value of $\theta_{23}$. We show, in particular, that in the ''nominal'' case of neutrino energy resolution $E_{res} = 30\%$ and zenith angle resolution $\theta_{zres} = 20^\circ$ and for, e.g., $\sin^2\theta_{23}=0.45~(0.58)$, HK can determine the average core density $\bar{\rho}_C$ at $2\sigma$ C.L. after 6500 days of operation with an uncertainty of (-14.5\%)/+39.5\% ((-9.3\%/+31.7\%). In the ''more favorable'' case of $E_{res}= 20\%$ and $\theta_{zres} = 10^\circ$, and if $\sin^2\theta_{23}=0.58~(0.45)$, the core density would be determined at $2\sigma$ C.L. with an uncertainty of (-8.3\%)/+9.8\% ((-9.2\%)/+11.3\%).

\end{abstract}


\end{titlepage}
\setcounter{footnote}{0}

\vspace{-0.4cm}

\newpage

\section{Introduction}
Our current knowledge 
of the density structure and composition of the Earth's interior is based 
primarily on 
seismological and geophysical data (see, e.g., 
\cite{McDonough:2003,Bolt:1991,Kennett:1998,Masters:2003,Nimmo:2022,Koper:2022,Waszek:2023,McDonough:2023,Noe:2023}).
These data were used to construct 
the Preliminary Reference Earth Model (PREM)  \cite{PREM} 
of the density distribution of the Earth. 
According to the PREM model, 
the Earth density distribution $\rho_{\rm E}$ is 
spherically symmetric, $\rho_{\rm E} = \rho_{\rm E}(r)$, 
$r$ being the distance from the Earth center,
and there are three major density structures or layers -
the inner core, outer core and the mantle, and
a certain number of substructures (shells).
There are also three thin ``surface'' layers just below
the Earth surface - the ocean and the two 
crust layers. The mantle has five shells in the model. 
The mean Earth radius is $R_{\oplus} = 6371$ km; the Earth 
inner core (IC) and outer core (OC) have radii 
$R_{IC} =  1221.5$ km and $R_{OC} =  3480$ km. 
The mantle layer is located between the outer core and 
the lower crust layer, i.e., for $r$ in the interval 
$R_{\rm C} \leq r \leq R_{\rm man}$, with $R_{\rm man} = 6346.6$ km.

 The determination of the radial density distributions in the mantle, 
outer core and inner core (or core)
$\rho_{\rm man}(r)$, $\rho_{\rm OC}(r)$ and $\rho_{\rm IC}(r)$
(or $\rho_{\rm C}(r)$) from seismological and geophysical data is not direct 
and suffers from uncertainties 
\cite{Bolt:1991,Kennett:1998,Masters:2003,Nimmo:2022}. 
It requires the knowledge, in particular, of the seismic wave speed 
velosity distributions in the interrior of the Earth, which depends 
on the pressure, tempertaure, composition and elastic properties 
of the Earth's interior that are not precisely known.  
Perhaps a rather conservative estimate of the uncertainty 
in the knowledge of $\rho_{man}(r)$ is $\sim 5\%$; for the core density  
$\rho_{c}(r)$ it is larger and can be noticeably 
larger \cite{Bolt:1991,Kennett:1998,Masters:2003,Nimmo:2022}. 

A precise knowledge of the Earth's density distribution 
and of the average densities 
of the mantle, outer core and inner core, or the core,
is crucial for understanding the physical conditions and
basic aspects of the structure and properties of the Earth's interior
(including the dynamics of mantle and core, the bulk composition of 
the Earth's three structures, the generation, properties and evolution 
of the Earth's magnetic field and the gravity field of the Earth) 
\cite{Bolt:1991,Yoder:1995,McDonough:2003,McDonough:2008zz}.

 Unique alternative methods of obtaining information about 
the density profile of the Earth are 
the neutrino  absorption and neutrino oscillation 
tomographies of the Earth. 
The propagation and oscillations of the active flavour neutrinos 
and antineutrinos $\nu_l$ and $\bar{\nu}_l$, $l=e,\mu,\tau$,
in the Earth are affected by the Earth matter.
The original idea of neutrino absorption Earth tomography is based on 
the observation that the cross section of the neutrino-nucleon 
interaction rises with energy. For neutrinos with energies 
$E_\nu \gtrsim$ a few TeV, the inelastic scattering off protons and neutrons 
leads to absorption of neutrinos and thus to 
attenuation of the initial neutrino flux. The magnitude of the 
attenuation depends on the Earth matter density along the neutrino path.
Attenuation data for neutrinos with different path-lengths 
in the Earth carry information about the matter density distribution 
in the Earth interior. The absorption method of Earth tomography 
with accelerator neutrino beams, which is difficult (if not impossible) 
to realise in practice was discussed first in 
\cite{PlaZavatt1973,VolkovaZatsepin1974} and later in grater detail in  
\cite{Nedyalkov:1981yy,Nedyalkov:2,DeRujula:1983ya,Wilson:1983an,Askar:1984,Borisov:1986sm,Borisov:1989kh,Winter:2006vg,Kuo95,Jain:1999kp,Reynoso:2004dt,Gonzalez-Garcia:2007wfs}.

  The oscillations between the active flavour neutrinos and antineutrinos, 
$\nu_l \leftrightarrow \nu_{l'}$ and 
 $\bar{\nu}_l \leftrightarrow \bar{\nu}_{l'}$, $l,l'=e,\mu$, 
having energies in the range $E \sim (0.1 - 15.0)$ GeV and 
traversing the Earth, can be strongly modified by the Earth matter effects 
(see, e.g., \cite{ParticleDataGroup:2018ovx}).
These modifications depend on the Earth matter density 
along the path of the neutrinos
\footnote{More precisely, they depend on 
the electron number density $N_e(r)$,
 $N_e(r) = \rho_{\rm E}(r) Y_e(r)/m_{N}$, $Y_e(r)$ and $m_N$ being 
the electron fraction number and nucleon mass.
It follows from the studies of the Earth matter composition
that $Y_e \cong (0.490-0.496)$ in the mantle and 
$Y_e \cong (0.466-0.471)$ in the core 
(see, e.g., \cite{Bardo:2015,Kaminski:2013,Sakamaki:2009,EarthRef}).
Taking into account these uncertainties has no effect on the 
results on the densities of the mantle and the core 
obtained in neutrino tomography studies (see, e.g.,\cite{Donini:2018tsg}).
The relation between $N_e(r)$ and $\rho_{\rm E}(r)$ 
can be used to obtain information about $Y_e(r)$, 
i.e., the composition of the core and mantle layers 
of the Earth (see, e.g., \cite{Rott:2015kwa,Bourret:2020zwg}). 
}. 
Thus, by studying the effects of Earth matter 
on the oscillations of, e.g., $\nu_\mu$ and $\nu_e$ ($\bar{\nu}_\mu$ and 
$\bar{\nu}_e$) neutrinos traversing the Earth along different trajectories 
it is possible to obtain information about the Earth density 
distribution. 

Atmospheric neutrinos (see, e.g., \cite{Gaisser:2002jj,Honda:2015fha}) 
are a perfect tool  for performing Earth tomography.
Consisting of  significant fluxes of 
muon and electron neutrinos and antineutrinos,
$\nu_\mu$, $\nu_e$, $\bar{\nu}_\mu$ and $\bar{\nu}_e$, 
produced in the interactions of cosmic rays with the Earth 
atmosphere, they have a wide range of energies 
spanning the interval from a few MeV to multi-GeV to multi-TeV. 
Being produced isotropically in the upper part of the Earth atmosphere 
at a height of $\sim 15$ km, they travel 
distances from $\sim 15$ km to 12742 km before reaching 
detectors located on the Earth surface, 
crossing the Earth along all possible directions 
and thus ``scanning'' the Earth interior.  
The interaction rates that allow to get 
information about the Earth density distribution 
can be obtained in the currently taking data IceCube
experiment \cite{IceCube,IceCube:Neutrino2024} and its 
planned upgrade \cite{IceCube:Neutrino2024}, in the 
ORCA \cite{KM3Net:2016zxf},   
Hyper Kamiokande \cite{Hyper-Kamiokande:2018ofw} 
and DUNE \cite{DUNE:2018tke} experiments which are under cosntruction,
and in the planned INO \cite{ICAL:2015stm} experiment.
 
In 2018 in \cite{Donini:2018tsg} the authors, following the  
idea of performing absorption neutrino tomography of the Earth 
with atmospheric neutrinos put forward in \cite{Gonzalez-Garcia:2007wfs},
used the data of the IceCube experiment on multi-TeV atmospheric  
$\nu_{\mu}$ and $\bar{\nu}_\mu$ with sufficiently long paths 
in the Earth \cite{IceCube:2016rnb}
and obtained information about the Earth density distribution. Their reults,
although not very  precise, broadly agree with the PREM model. 
In particular, they contain evidence at $2\sigma$ C.L. 
that the core is denser than the mantle.
The authors used the IceCube data to ``weight'' the Earth with neutrinos:
the value of the Earth mass  found in 
\cite{Donini:2018tsg}, $M^\nu_\oplus = (6.0 ^{+1.6}_{-1.3})\times 10^{24}$ kg, 
is in good agreement  with gravitationally determined value  
\cite{EarthMI1,EarthMI2}, 
$M_\oplus = (5.9722 \pm 0.0006) \times 10^{24}~{\rm kg}$.
%
This was the first time 
the study of neutrinos traversing the Earth provided information 
of the Earth interior and marked the beginning of 
real experimental data driven neutrino tomography of the Earth.

The neutrino oscillation tomography of the Earth with 
IceCube, ORCA, DUNE and INO detectors is discussed in 
\cite{Agarwalla:2012uj,Rott:2015kwa,Winter:2015zwx,Bourret:2017tkw,Bourret:2020zwg,Kumar:2021faw,Capozzi:2021hkl,Kelly:2021jfs,Denton:2021rgt,Maderer:2022toi,Upadhyay:2022jfd,Raikwal:2023jkf,Upadhyay:2024gra} 
(the early studies include \cite{Choubey:2008_2011,Choubey:2014} 
briefly described in \cite{Capozzi:2021hkl}; see also 
\cite{Ohlsson:2001ck}).

In the present article we investigate the possibility 
to get information about the Earth density structure 
performing neutrino tomography of the Earth 
with the Hyper-Kamiokande (HK) experiment by studying the oscillations 
of atmospheric neutrinos (antineutrinos)  $\nu_\mu$ and $\nu_e$ 
($\bar{\nu}_\mu$ and $\bar{\nu}_e$) with different energies and pathlengths
in the Earth. In \cite{Hyper-Kamiokande:2018ofw} the HK collaboration 
presented results on the potential of the HK experiment 
to determine the composition (i.e., the Z/A ratio) of the Earth core 
by using neutrino oscillation tomography of the Earth.  The Hyper-Kamiokande detector is under construction 
and is foreseen to begin to take data in 2027.

The article is organised as follows. In Section 2 
we review basic elementary particle and 
geophysics aspects of the study we perform. 
In particular, in subsection 2.1
we discuss in detail the implementation 
of the Earth total mass, moment of inertia and hydrostatic 
equilibrium constraints in our analysis.
Section 3 is devoted to the analysis methodology
(including discussions of HK simulation, 
event selection, the employed atmospheric neutrino fluxes, 
oscillation probabilities evaluation, 
enhanced datasets generation and sensitivity calculation).
In Section 4 we present the results of our analysis 
of HK sensitivity to the Earth density structure.
Section 5 contains our conclusions.

%
\section{PREM Input, Oscillation Probabilities 
and Earth Data Constraints}
\label{sec:2}
%
%
 We use the PREM model as a reference model of the 
Earth density distribution $\rho_{\rm E}(r)$ 
and  assume that the location of the 
mantle-core and the outer core-inner core  
boundaries are correctly described by the model. 
In the PREM model, as we have already briefly discussed 
in the Introduction, the Earth density distribution 
$\rho_{\rm E}$ is assumed to be 
spherically symmetric, $\rho_{\rm E} = \rho_{\rm E}(r)$, 
$r$ being the distance from the Earth center.
The mean Earth radius is $R_{\oplus} = 6371$ km; 
the Earth core (C) has a radius of $R_C = 3480$ km 
with the inner core (IC) and outer core (OC) extending respectively 
from $r = 0$ to $r = R_{IC} = 1221.5$ km, 
and from $r = 1221.5$ km to $r = R_{OC} = R_C = 3480$ km.
The mean densities of the mantle, 
IC, OC and core are respectively $\bar{\rho}_{\rm man} = 4.66$ g/cm$^3$,
 $\bar{\rho}_{\rm IC} = 12.89$ g/cm$^3$,
$\bar{\rho}_{\rm OC} = 10.90$ g/cm$^3$ and 
$\bar{\rho}_{\rm C} = 10.99$ g/cm$^3$.
If we include the three ``surface'' layers 
in the mantle, as we are going to do in what follows, 
the mantle extends from $r=R_C = 3480$ km to $r = R_{\oplus} = 6371$ km
and the mean density of the mantle changes 
to $\bar{\rho}_{\rm manco} = 4.45$ g/cm$^3$ 
(see, e.g., \cite{Petcov:2024icq}).
The change of density between the mantle and the outer core 
is described by a step function.

For a spherically symmetric Earth density
distribution, the neutrino trajectory
in the Earth is specified by the value
of the zenith angle $\theta_z$ of the trajectory.
For $ 90^\circ \leq \theta_z\leq (180^\circ - 33.17^\circ)$,
or path lengths $L \geq 10665.7$ km,
neutrinos cross the Earth core.
The path length for neutrinos which
cross only the Earth mantle is given
by $L =  -\,2 R_{\oplus}\cos \theta_z$.
If neutrinos cross the Earth core,
the lengths of the paths in the mantle,
$2L^{\rm man}$, and in the core,
$L^{\rm core}$, are determined by:
$L^{\rm man} = -\, R_{\oplus}\cos \theta_z -
(R_c^2 - R^2_{\oplus}\sin^2\theta_z)^{1/2}$,
$L^{\rm core} = 2(R_c^2 - R^2_{\oplus}\sin^2\theta_z)^{1/2} $,
with $(180^\circ - 33.17^\circ) \leq \theta_z \leq 180^\circ$.
Correspondingly, the neutrinos cross the core, the inner core
and the outer core for $\cos \theta_z$ lying respectively 
in the intervals [-1.00, -0.84], [-1.00,-0.98] and [-0.98,-0.84] 
%
%
%

In contrast to seismic waves, which are usually 
reflected or refracted at density jumps, neutrino oscillations are 
essentially not sensitive to changes of density 
in the Earth mantle and core taking place over distances 
which are smaller than the neutrino oscillation length 
\cite{Ohlsson:2001ck} (see also \cite{ParticleDataGroup:2018ovx})
that in the cases we are going to study 
is typically $\sim 500$ km 
\footnote{The estimate we quote refers to the relevant quantity, 
$L_{\rm osc}/(2\pi)$, $L_{\rm osc}$ being the neutrino oscillation length.
}. 
Thus, they are sensitive, in general, to the average densities of 
the mantle (or layers of the mantle having width $\sim 1000$ km),
and of the inner core  and the outer core
(or possibly of two layers of the outer core (mantle) 
each having width $\sim 1000$ km). 
However, through the mantle-core interference 
(or neutrino oscillation length resonance- (NOLR-) like) effect 
\cite{Petcov:1998su,Chizhov:1999az,Chizhov:1999he}
(see also \cite{Chizhov:1998ug,AMS:2006hb,AMS:2005yj} 
and references quoted therein), 
the neutrino oscillations are sensitive 
to densities of the mantle and the 
core as well as to the difference of the  densities 
of the mantle and the core, i.e., to the magnitude 
of the density ``jump'' in the mantle-core 
narrow transition zone. 
They might be sensitive, in principle, also 
to the difference between the IC and OC (average) densities.
 
 The Earth matter effects in the neutrino oscillations of interest 
depend on the matter potential 
\cite{MSW1,Barger80,Langa83}
\begin{equation}
V = \sqrt{2}\,G_{\rm F}\, N_e\,,
\label{eq:V}
\end{equation}
%
which involves the electron number density 
$N_e$ along the path of the neutrinos.
The relation between the Earth density and 
electron number density includes the electron 
fraction number $Y_e$ (or $Z/A$ factor) 
of the corresponding Earth 
structure or layer: $N^{(E)}_{e}(r) = \rho_{\rm E}(r)\,Y_e/m_{\rm N}$, 
where $m_{\rm N}$ is the nucleon mass. 
For isotopically symmetric matter 
$Y_e = 0.5$. However, the compositions of the 
Earth mantle and core are not exactly 
isotopically symmetric.
For the outer core, for example, 
different composition models give a value 
of $Y_e$ in the interval $Y_e^{oc} = 0.466 - 0.471$ 
(see, e.g., 
\cite{Bardo:2015,Kaminski:2013,Sakamaki:2009,McDonough:2003,EarthRef}).
The value of $Y_e$ in the mantle is closer to 0.5 
\cite{PREM,McDonough:2003}:
 $Y_e^{man} = 0.490 - 0.497$. 
In this study we will use the following default values of 
$Y_e$ in the mantle and the core: 
$Y_e^{man} = 0.50$ and  $Y_e^{c} = 0.50$.
The results we will obtain 
change insignificantly if instead we used 
values of $Y_e^{oc}$ and  $Y_e^{man}$ 
from the above  quoted intervals. 

 As is well known, for NO (IO) neutrino mass spectrum, 
the matter effects can lead to strong enhancement of the neutrino 
(antineutrino) transition probabilities of interest 
$P(\nu_l \rightarrow \nu_l') \equiv P_{ll'}$ and 
$P(\bar{\nu}_l \rightarrow \bar{\nu}_l') \equiv \bar{P}_{ll'}$, 
$l\neq l' = e,\mu$, and $l = e$, $l' = \tau$,
for neutrino energies  $E\sim (6-10)$ GeV and $\sim (3-5)$ GeV, 
corresponding respectively to the resonance 
in the mantle \cite{MSW1,MSW2} (see also \cite{Barger80})
and to the mantle-core interference 
(NOLR) effect \cite{Petcov:1998su}. In our analysis we consider all neutrino interactions in the energy interval, $E = (1-100)$ GeV, well covering the most sensitive range to density dependent effects which are contained in the energy interval $E\sim (2-10)$ GeV, and taking into account the effects of the energy resolution.

The neutrino oscillation probabilities are calculated 
taking into account the variation of densities in inner core, outer core, 
and the two mantle layers, so that they reproduce 
the results corresponding to 
the variations predicted by the PREM model.  
A more detail discussion is given in subsection 3.4.
The neutrino oscillations parameters used as input in the calculations 
of the probabilities are given in Table 1. At neutrino energies 
$E_\nu \geq 2$ GeV we will be interested in the  
effects of $\Delta m^2_{21}$ and the Dirac CP-violation phase 
$\delta$ in the oscillation probabilities 
are subleading being strongly suppressed.
We do not take into account the uncertainties in the 
determination of the neutrino oscillation parameters 
since their effect is much smaller than the effects of 
the systematic and statisctical uncertaties accounted for in the 
calculations.
\begin{table}[htbp]
\centering
\begin{tabular}{|c|c|c|c|c|c|c|}
\hline\hline
Ordering & $s^2_{13}$  & $s^2_{12}$ & $s^2_{23}$  & $\delta_{CP}$  & $\Delta m^2_{32}$ ($10^{-3}$ eV$^2$) &  $\Delta m^2_{21}$ ($10^{-5}$ eV$^2$)  \\
\hline\hline
Normal & 0.022 & 0.307 & 0.45 & -1.75 &  2.40 & 7.53\\
\hline\hline
\end{tabular}
\caption{PMNS parameters used in the study,
$s^2_{ij} \equiv \sin^2\theta_{ij}$.
The values correspond to the best fit point in 
Ref.~\cite{Super-Kamiokande:2023ahc} for normal 
neutrino mass ordering.
}
\label{tab:sk_bfp}
\end{table}
%

%
\subsection{Earth Total Mass, Moment of Inertia and Hydrostatic 
Equilibrium Constraints}
\label{EMIHEC}
%

In order to estimate the sensitivity of the HK detector 
to possible deviations of the core (IC + OC) 
and the two mantle layers densities from those obtained 
using geophysical and seismological data and described by
PREM, in the statistical analysis we have performed 
we varied the density of each of these  structures, 
as is typically done in the neutrino tomography studies 
of the Earth interior.  We will assume that 
that the  variations and changes of 
$\rho_{\rm IC}$, $\rho_{\rm OC}$, 
 $\rho_{\rm lman}$ and   $\rho_{\rm uman}$ 
are by constant $r$-independent factors 
$\kappa_{\rm IC}$, $\kappa_{\rm OC}$,  $\kappa_{\rm lman}$ 
and  $\kappa_{\rm uman}$:
\begin{equation}
\rho^\prime_{i} = (1 + \kappa_i)\,\rho_{i}\,,~i={\rm IC,OC,lman,uman}\,, 
\kappa_i > -\,1\,.
\label{eq:kappai4}
\end{equation}
%
In this study we will consider the case of 
variation of $\rho_{\rm IC}$ and $\rho_{\rm OC}$ 
by the same factor:
\begin{equation}
\kappa_{IC}  = \kappa_{OC} = \kappa_{C}\,. 
\label{eq:kappaICOC}
\end{equation}
%

 In our analysis we implemented the constraints on the variations of 
$\rho_{IC}$, $\rho_{OC}$, $\rho_{lman}$ and $\rho_{uman}$ 
following from the precise knowledge of the 
Earth mass and moment of inertia:
\cite{EarthMI1,EarthMI2,EarthMI3}:
\begin{eqnarray}
\label{eq:ME}
& M_\oplus =  (5.9722 \pm 0.0006)\times 10^{24}~{\rm kg}\,,\\[0.25cm]
\label{eq:IE}
&  I_\oplus = (8.01736 \pm 0.00097)\times 10^{37}~{\rm kg\, m^2}\,.
\end{eqnarray}
%
The mean value of $I_\oplus$ can be related to the Earth mean total mass and 
radius $R_\oplus$:
\begin{equation}
I_\oplus = 
C_{\rm G(PREM)} \,M_\oplus\,R^2_\oplus\,.
\label{eq:IMR2}
\end{equation}
%
where $C_{\rm G}$  ($C_{\rm PREM}$) 
is a constant determined in gravity experiments 
(the PREM model):
\begin{equation}
C_{\rm G} = 0.330745\,,~~~C_{\rm PREM} = 0.33080\,.
\label{eq:C}
\end{equation}
%

It is well known from the seismological and geophysical studies that in
order for Earth to be in hydrostatic equilibrium  
the following inequalities should always hold:
\begin{equation}
\rho^\prime_{man} < \rho^\prime_{OC} <  \rho^\prime_{IC}\,.
\label{eq:EHE1}
\end{equation}
%
In the present analysis we will require that these Earth hydrostatic
equilibrium (EHE) conditions are satisfied by the average densities 
of the corresponding Earth layers, 
$\bar{\rho}^\prime_i = \bar{\rho}_i (1 + \kappa_i)$.

 We implement the total Earth mass $M_\oplus$ and moment of ineria $I_\oplus$
constraints by compensating the variation of the density in the structure
of interest, for example,  $\rho_{OC}$ or  $\rho_{C}$,
by corresponding change of the densities in two   
other density structures such that the EHE
conditions are also satisfied.
\begin{figure*}[htpb!]
\centering
\includegraphics[width=0.45\textwidth]{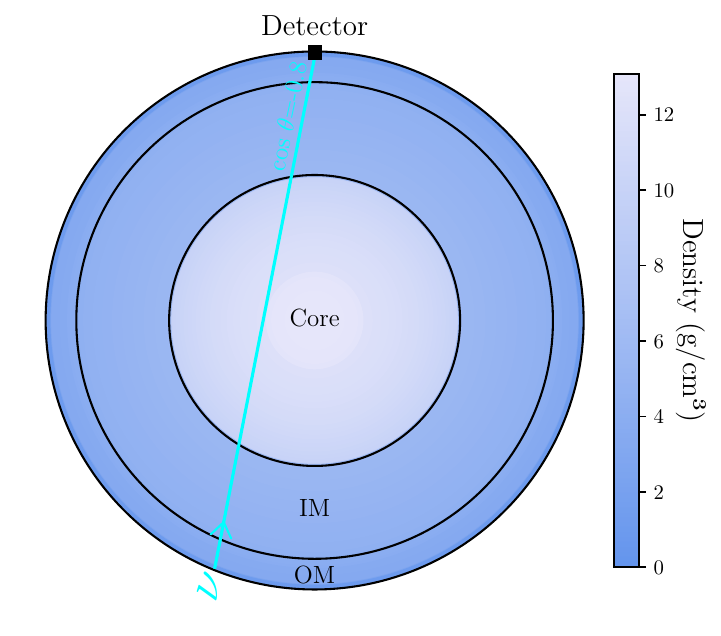}
\caption{A diagram of the Earth interior under study. 
The angle $\cos \theta = -1$ corresponds to an up-going neutrino crossing 
the Earth along the diameter before being detected.
}
\label{fig:3lmodel}
\end{figure*}
%

It was shown in \cite{Petcov:2024icq} that if we approximate
the Earth density structure with the standard three layers -  
the inner core, outer core and mantle,
the $M_\oplus$, $I_\oplus$ and EHE constraints
limit sverely the allowed ranges of variations of densities
in these layers. For example, $\kappa_{OC}$
is constrained to lie in the interval $-\,0.036 \leq \kappa_{OC} < 0.01$.
None of the currently operating and planned future detectors 
can be sentitive to such small deviations of
$\bar{\rho}_{OC}$, $\bar{\rho}_{IC}$  and$ \bar{\rho}_{man}$  from thier
respective PREM values. 

 In our analysis we will use effectively a three-layer model 
of Earth density distribution consisting in  the core and
two - lower and upper - mantle layers
\footnote{The same three-layer Earth density structure was used in the
study of the DUNE detector potential to perform neutrino
tomography of the Earth \cite{Kelly:2021jfs}.
}.
The lower mantle layer 
extends from $R_c = 3480$ km to $R_{\rm lman} = 5701$ km,  
and has an average density  $\bar{\rho}_{\rm lman} = 4.90$ g/cm$^3$ , 
while the upper mantle one extends from  $R_{\rm lman} = 5701$ km to 
$R_{\rm man} = 6371$ km, having an average density
$\bar{\rho}_{\rm uman} = 3.60$ g/cm$^3$. 
This density structure is shown in Fig. \ref{fig:3lmodel}.

Since in the case we study the mantle is devided into two layers  
and $\kappa_{IC}  = \kappa_{OC} = \kappa_{C}$,
so that $\bar{\rho}^\prime_{OC} <  \bar{\rho}^\prime_{IC}$
is always fulfilled, we consider the following form
of the EHE conditions:
\begin{equation}
\bar{\rho}^\prime_{uman} < \bar{\rho}^\prime_{lman} < \bar{\rho}^\prime_{C}\,.
\label{eq:EHE2}
\end{equation}
%

\begin{figure*}[htpb!]
\centering
\includegraphics[width=0.5\textwidth]{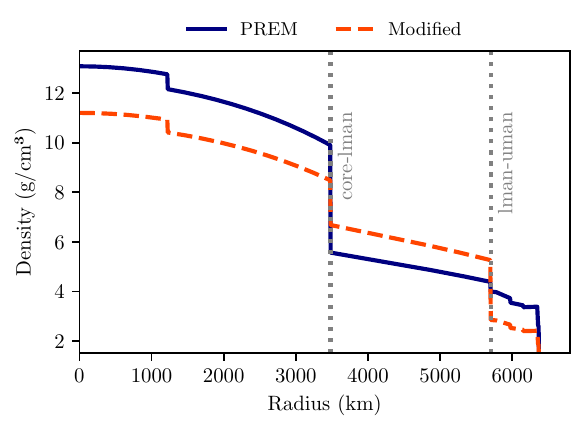}
\includegraphics[width=0.45\textwidth]{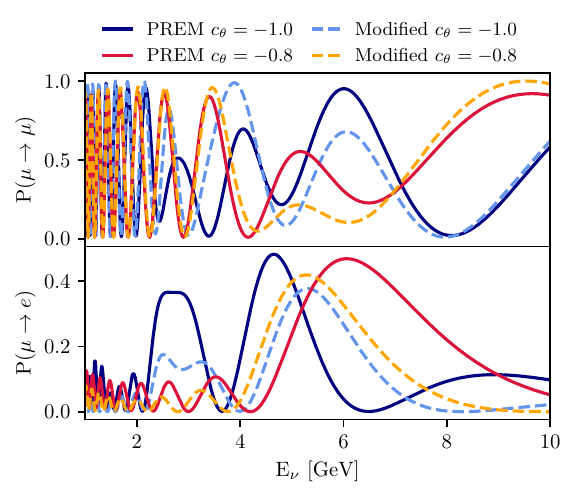}
\caption{Left panel: 
Earth density according to PREM and an example solution for a modified Earth 
with fixed
$\kappa_{lman} = 0.2$.
Right panel: $\nu_\nu$ disappearance and $\nu_e$ appearance probabilities for 
two values of the azimuth angle and 
the Earth models shown in the left panel.
}
\label{fig:TomographyConcept}
\end{figure*}

In the specific case we consider  
with the EHE constrints given in Eq. (\ref{eq:EHE2}),
we implement the $M_\oplus$ and $I_\oplus$ 
constraints by compensating the variation of the density of the
core $\bar{\rho}_{C}$ by corresponding change of the densities in two   
mantle layers $\bar{\rho}_{lman}$ and $\bar{\rho}_{uman}$.
This is illustrated in the left panel in Fig. \ref{fig:TomographyConcept}.
A given modification of the PREM density profile of the Earth 
leads to a corresponding modification of neutrino oscillation probabilities.
For the case of the densities shown in the left panel  
in Fig. \ref{fig:TomographyConcept} 
we show in the right panel of the same figure 
the changes of the oscillation probabilities for 
$\cos\theta_z = -\,1,\,-\,0.8$.
%

We follow the procedure described in \cite{Petcov:2024icq}.
The values of $M_\oplus$ and $MoI \equiv I_\oplus$  are calculated by
integrating the polynomials describing the density 
distributions in the PREM model. 
Using the $M_\oplus$ and $I_\oplus$ constraints we can express 
the constants parametrising the deviations of the lower and upper mantle 
average densities from their PREM values,
$\kappa_{lman}$ and $\kappa_{uman}$. 
in terms of the constant $\kappa_{C}$ parametrising the deviation 
of the core  (outer core) average density. 
 Working with $\bar{\rho}^\prime_{uman}$, $\bar{\rho}^\prime_{lman}$ and 
$\bar{\rho}^\prime_{C}$ we find:
\begin{equation}
\kappa_{lman} = f_{lman}\,\kappa_{C}\,,~f_{lman} = -\,1.43;
\kappa_{uman} = f_{uman}\,\kappa_{C}\,,~f_{uman} = 2.070\,.
\label{eq:kmanvskc}
\end{equation}
%
We remark that if we use 
 $\bar{\rho}^\prime_{OC}$ instead of $\bar{\rho}^\prime_{C}$ 
we obtain somewhat different values of $f_{lman}$ and $f_{uman}$:        
$f_{lman} = -\,1.539$ and $f_{uman} = 2.425$.
Thus, in both cases in order to compensate the variation of the density in the 
core, the density should be increased in one of the mantle 
layers and decreased in the second mantle layer.

In the case of $\kappa_{C} < 0$, the relevant  
EHE constraint in Eq. (\ref{eq:EHE2})  is  
$\bar{\rho}_{lman}(1 + \kappa_{lman}) < \bar{\rho}_{C}(1 + \kappa_C)$.
Using the relation between $\kappa_{lman}$ and $\kappa_C$
in Eq. (\ref{eq:kmanvskc}) and the values of 
$\bar{\rho}_{lman}$ and $\bar{\rho}_{C}$,
one can show this constraint is satisfied 
for $\kappa_{C} > -\,0.335$.
This in turn implies that 
$\kappa_{lman} < 0.482$ and $\kappa_{uman} > -\, 0.693$.
Summarising, in the case of  $\kappa_{C} < 0$, the 
 $M_\oplus$, $I_\oplus$ and the
EHE constraints  (\ref{eq:EHE2}) lead to:
 \begin{equation}
\kappa_{C} > -\,0.335\,,~\kappa_{lman} < 0.482\,,
~\kappa_{uman} > -\,0.693~~
(\kappa_{C} <0)\,.
 \label{eq:kcNeg}
 \end{equation}
%

The most  stringent constraint on $\kappa_{C} > 0$ follows 
from the requirement that when we vary $\kappa_{C}$, and correspondingly 
$\kappa_{uman}$ and  $\kappa_{lman}$, the inequality 
$\bar{\rho}^\prime_{uman} < \bar{\rho}^\prime_{lman}$ should always hold 
(Eq. (\ref{eq:EHE2})). This condition reads: 
$\bar{\rho}_{uman}( 1 + \kappa_{uman}) < \bar{\rho}_{lman}( 1 + \kappa_{lman})$.
Employing the relations between $\kappa_{uman}$, $\kappa_{uman}$ 
and $\kappa_{C}$ given in Eq. (\ref{eq:kmanvskc}) we get:
\begin{equation}
\kappa_{C} <
\dfrac{\bar{\rho}_{lman} - \bar{\rho}_{uman}}
{f_{lman}\bar{\rho}_{lman} + f_{uman}\bar{\rho}_{uman}}  \cong 0.089\,, 
\label{eq:kcNeg2}
\end{equation}
%
where we have used the numerical values of 
$f_{lman}$, $\bar{\rho}_{lman}$, $f_{uman}$ and $\bar{\rho}_{uman}$.
This upper limit on $\kappa_{C} > 0$ in turn implies
$\kappa_{uman} < 0.184$ and $\kappa_{lman} > -\,0.128$.

Combing the results for negative and positive $\kappa_{C}$, 
which correspond to the EHE constraints in Eq. (\ref{eq:EHE2})
for the average densities, we have:
\begin{align}
 &-\,0.33< \kappa_{C} < 0.09 \nonumber\\
    &-\,0.13 < \kappa_{lman} < 0.48 \nonumber\\
  & -\,0.69 < \kappa_{uman} < 0.18\,.
 \label{eq:ka3l}     
\end{align}
%

We get similar ranges of allowed values of $\kappa_{C}$,
$\kappa_{lman}$ and $\kappa_{lman}$ if in the preceding discussion we
used $\bar{\rho}^\prime_{OC}$ instead of  $\bar{\rho}^\prime_{C}$.
At the same time, using the EHE constraints in Eq. (\ref{eq:EHE1})
leads to weaker restrictions on $\kappa_{C}$,
$\kappa_{lman}$ and $\kappa_{lman}$
\footnote{This case was analised in Ref. \cite{Petcov:2024icq} 
  with $\bar{\rho}_{IC}$ kept at its PREM value and the mantle
  devided in upper and lower mantle layers.
We note that in the expression of 
the denominator in the right-hand side of the inequality 
in  Eq. (69) in \cite{Petcov:2024icq}, the signs in the front of the 
second and third terms should be negative, i.e., the correct form of 
this expression is:
$\bar{\rho}_{ OC} -\,\bar{\rho}_{lman}\,C_{lman}\,f_{lman} -
\,\bar{\rho}_{uman}\,C_{uman}\,f_{uman}$, where 
$f_{lman}$, $f_{uman}$ and 
$C_{lman}$ and $C_{uman}$ 
are defined in Eqs. (65) and (68), respectively. 
Correspondingly, the inequality in
Eq. (70) changes to $\kappa_c > -\,0.496$.
However, the strongest lower limit in the case of  $\kappa_c < 0$ 
follows from the EHE constraint 
$\bar{\rho}_{uman}(1 + \kappa_{uman}) < \bar{\rho}_{OC}(1 + \kappa_{OC})$:
 $\kappa_c > -\,0.334$.
The strongest upper limit on
$\kappa_c > 0$ follows from the EHE condition that
$\bar{\rho}_{uman}(1 + \kappa_{uman}) < \bar{\rho}_{lman}(1 + \kappa_{lman})$: 
$\kappa_c < 0.086$.
Thus, Eq. (71)  in \cite{Petcov:2024icq} should read: 
$-\,0.334 < \kappa_c < 0.086$. 
}.

 We have performed also a numerical scan to determine 
the values of  $\kappa_i$, $i=C,lman,uman$,
which satisfy the $M_\oplus$, $I_\oplus$ and the EHE 
constraints. In this scan the conservation 
of $M_\oplus$ and $I_\oplus$ are required to be  fulfilled with 
a numerical uncertainty of $1\%$ 
of the nominal values of $M_\oplus$ and $I_\oplus$.
The EHE constraint is imposed by requiring that the inequalities 
in Eq. (\ref{eq:EHE2}) are always satisfied for the corresponding 
average densities. Following the described  procedure we find:
\begin{align}
 &-\,0.30 < \kappa_{C} < 0.10 \nonumber\\
    &-\,0.15 < \kappa_{LMT} < 0.45 \nonumber\\
    &-0.65 < \kappa_{UMT} < 0.10
\label{eq:ka3lnum}     
\end{align}
%
The above results obtained numerically we are going to use in 
what follows agree vey well with the results in Eq. (\ref{eq:ka3l}) 
derived analytically. 

It follows from Eqs. (\ref{eq:ka3l}) and (\ref{eq:ka3lnum}) that 
the  $M_\oplus$, $I_\oplus$ and EHE constraints limit 
in the case of the three-layer Earth density structure we are considering 
the possible deviations of  $\bar{\rho}_{C}$, $\bar{\rho}_{lman}$ and 
$\bar{\rho}_{uman}$ from their respective 
PREM values of 
10.99 g/cm$^3$, 4.90 g/cm$^3$ and 3.60 g/cm$^3$ 
to the following rather narrow intervals: 
$7.36~{\rm g/cm^2}  < \bar{\rho}_{C} < 11.98$  g/cm$^3$,
$4.26~{\rm g/cm^3} < \bar{\rho}_{lman} < 7.25 $ g/cm$^3$ 
and  $1.12~{\rm g/cm^3} < \bar{\rho}_{lman} < 4.23$ g/cm$^3$, 
where 
\footnote{Similar restrictions were obtained in 
Ref. \cite{Kelly:2021jfs}. Numerically, our results differ 
somewhat from those found in \cite{Kelly:2021jfs} 
since in the constraint following from the conservation 
of $I_\oplus$ the authors used the average densities 
of the considered three density layers 
present in the expression for $M_\oplus$  
(Eqs. (5.3) and (5.4) in \cite{Kelly:2021jfs}), while, 
we have utilised the results of the analysis 
in \cite{Petcov:2024icq} which showed 
that the average densities in the expression 
for $I_\oplus$ differ somewhat form those in the expression 
for  $M_\oplus$.}
we have used the ranges of $\kappa_i$ given in Eq. (\ref{eq:ka3l}).

%
\section{Analysis Methodology}
%


 This study uses the public release~\cite{SuperKamiokande2023} of the 
simulated atmospheric neutrino event rate predictions generated by the 
Super-Kamiokande (SK) collaboration in its latest  atmospheric neutrino 
oscillation analysis~\cite{Super-Kamiokande:2023ahc}. It includes realistic 
flux~\cite{Honda:2011nf}, neutrino interaction modeling using NEUT~\cite{Hayato:2009zz, Hayato:2021heg} , detector effects and analysis 
sample selection criteria, see Sec.~\ref{sksamples}. Based on those 
simulations, we extrapolate the Earth tomography sensitivity to 
Hyper-Kamiokande (HK), a detector eight times larger than SK, but based on 
the same detection concept and with small instrumentation differences. 
Additionally, we modify the nominal distributions, as later discussed in 
Sec.~\ref{fds_generation}, to assess the potential impact of enhanced 
energy and/or angular resolutions in HK. The reported sensitivities 
are obtained as described in Sec.~\ref{chi2calc}.

%
\subsection{SK simulation input}
%

Our study utilizes as input the simulation release in 
Ref.~\cite{SuperKamiokande2023}. It consists of the expected event rates, 
and one-dimensional true energy and angle event distributions for a 
list of 931 angle and energy bins divided in 29 samples, corresponding to 
29 different sets of event selection conditions. The distributions are 
released as 1D true neutrino energy and true azimuth angle quantiles 
at 2.3\%, 15.9\%, 50\%, 84.1\% and 97.7\% of the distribution. The predictions 
are divided in three major blocks, consisting of the unoscillated, and 
oscillated predictions assuming either the normal or the inverted mass 
ordering. For each, the predictions are divided in six components: 
$\nu_e$, $\nu_\mu$, $\bar{\nu}_e$, $\bar{\nu}_\mu$, $\nu_\tau + \bar{\nu}_\tau$ 
and neutral current (NC). Additionally, the measured data counts in every 
bin are provided. 
 It is important to note that 
it is not possible to use 
data counts 
as Super-Kamiokande does only release the prefit event rate 
expectations, namely, without applying the systematic pulls at the best 
fit point. Therefore, we focus in this study on the analysis of the nominal 
simulation expectations, which are in any case an excellent proxy to study 
the sensitivity of future similar experiments.

%
\subsection{Selection Samples}
\label{sksamples}
%
\begin{figure*}[hpb!]
\centering
\includegraphics[width=0.32\textwidth]{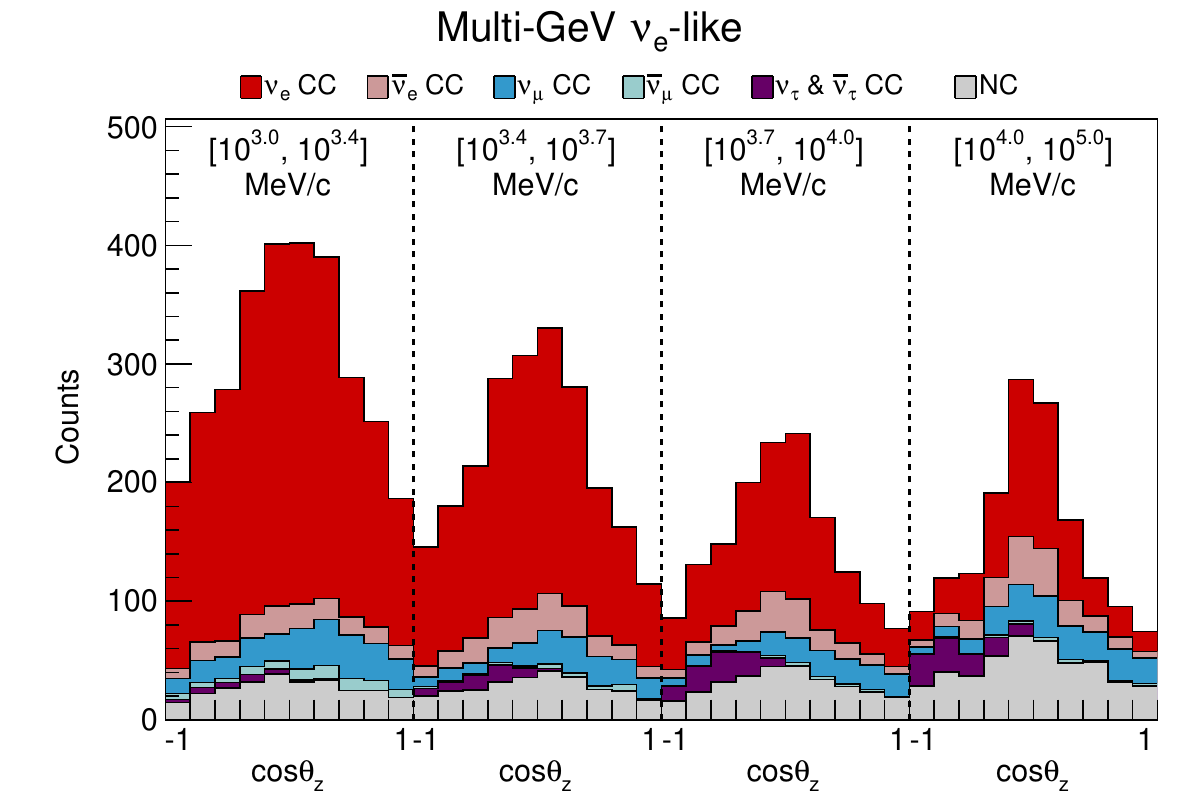}
\includegraphics[width=0.32\textwidth]{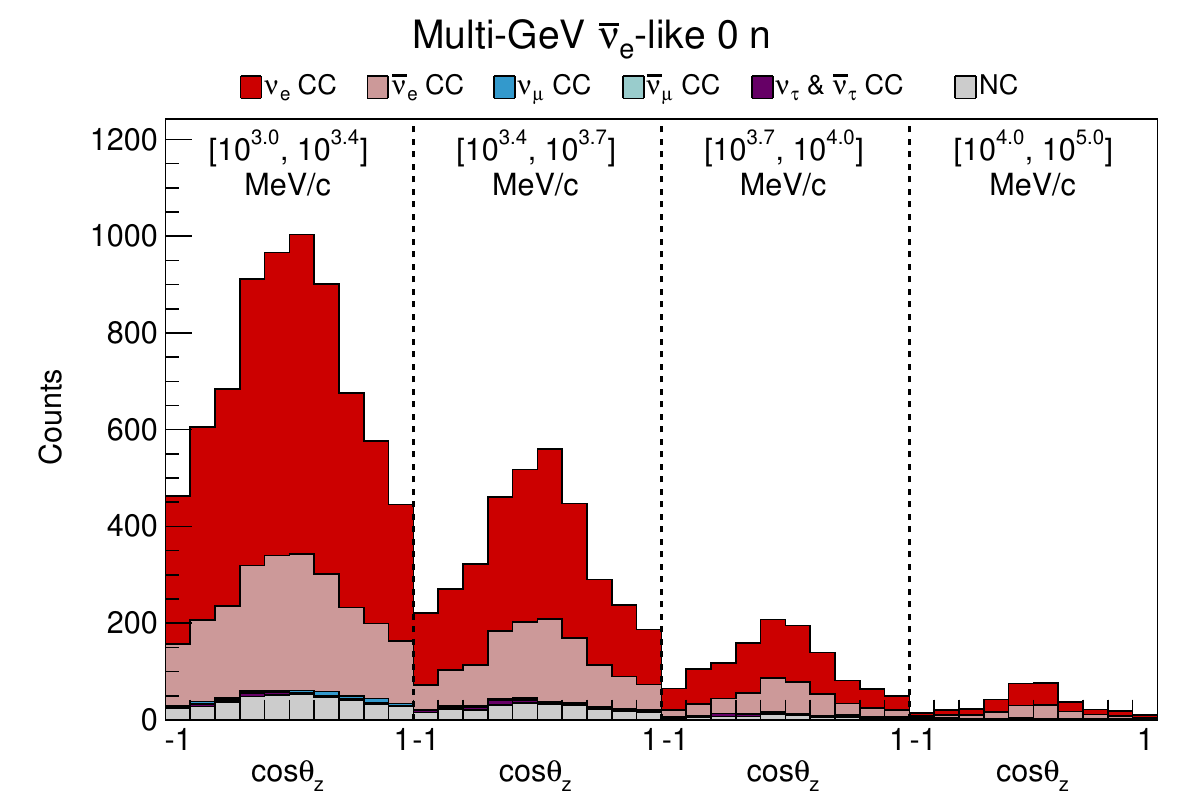}
\includegraphics[width=0.32\textwidth]{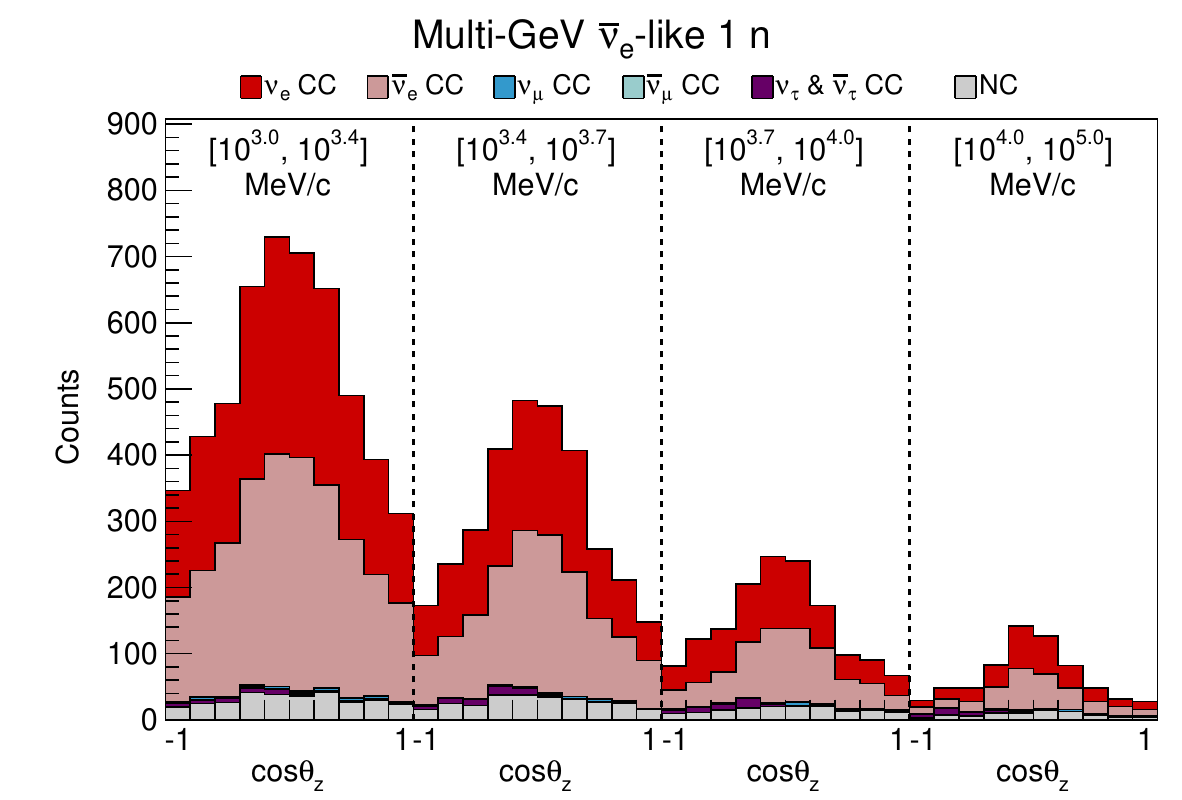}
\includegraphics[width=0.32\textwidth]{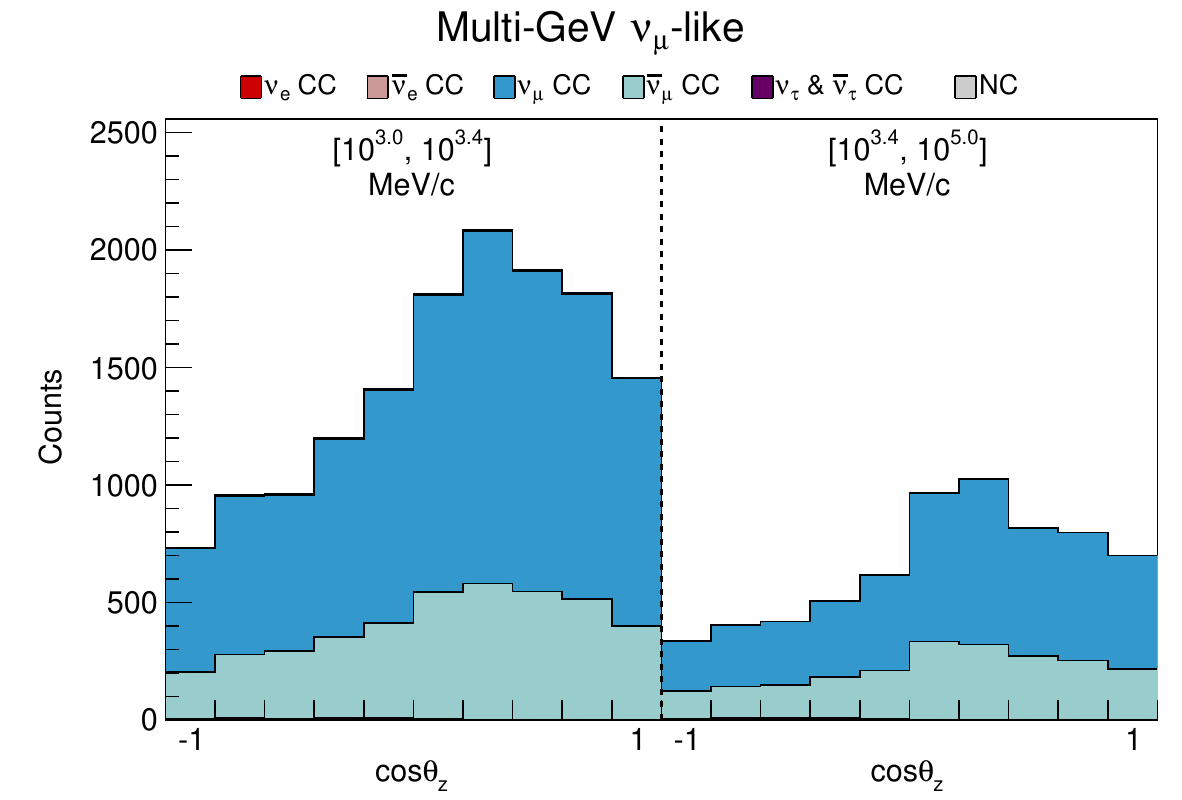}
\includegraphics[width=0.32\textwidth]{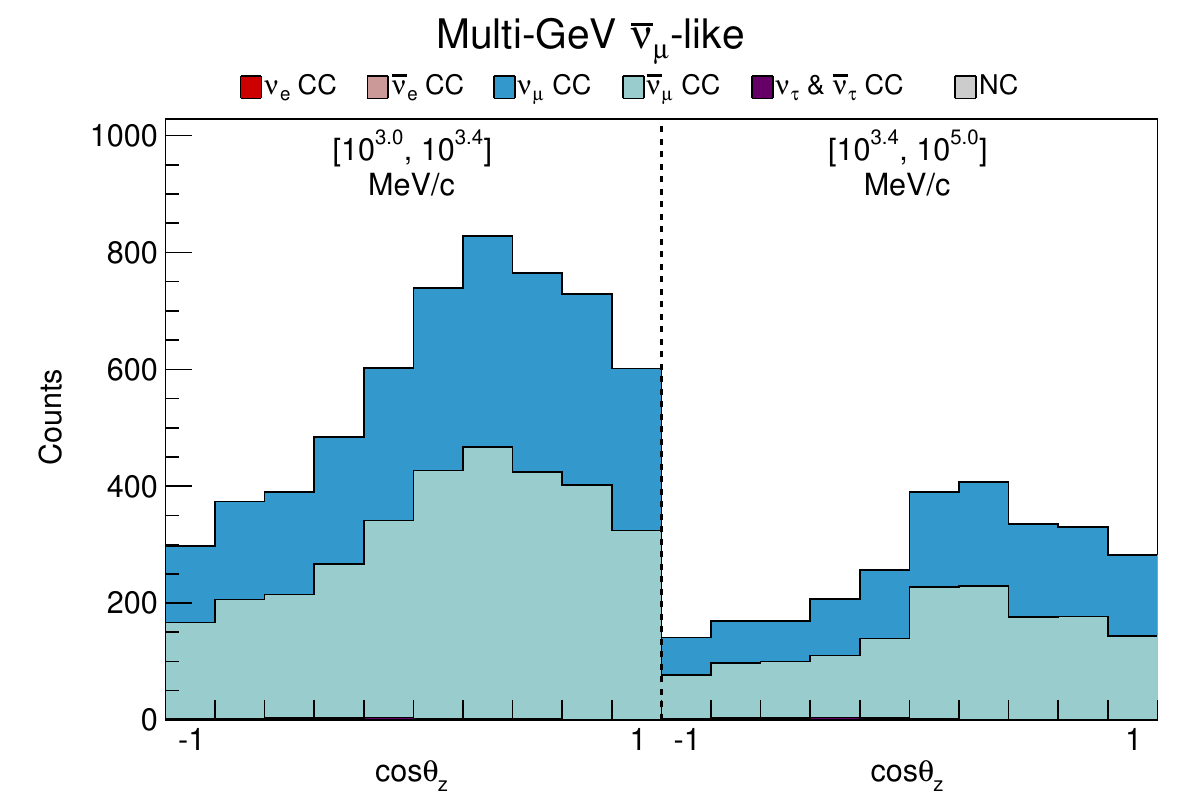}
\includegraphics[width=0.32\textwidth]{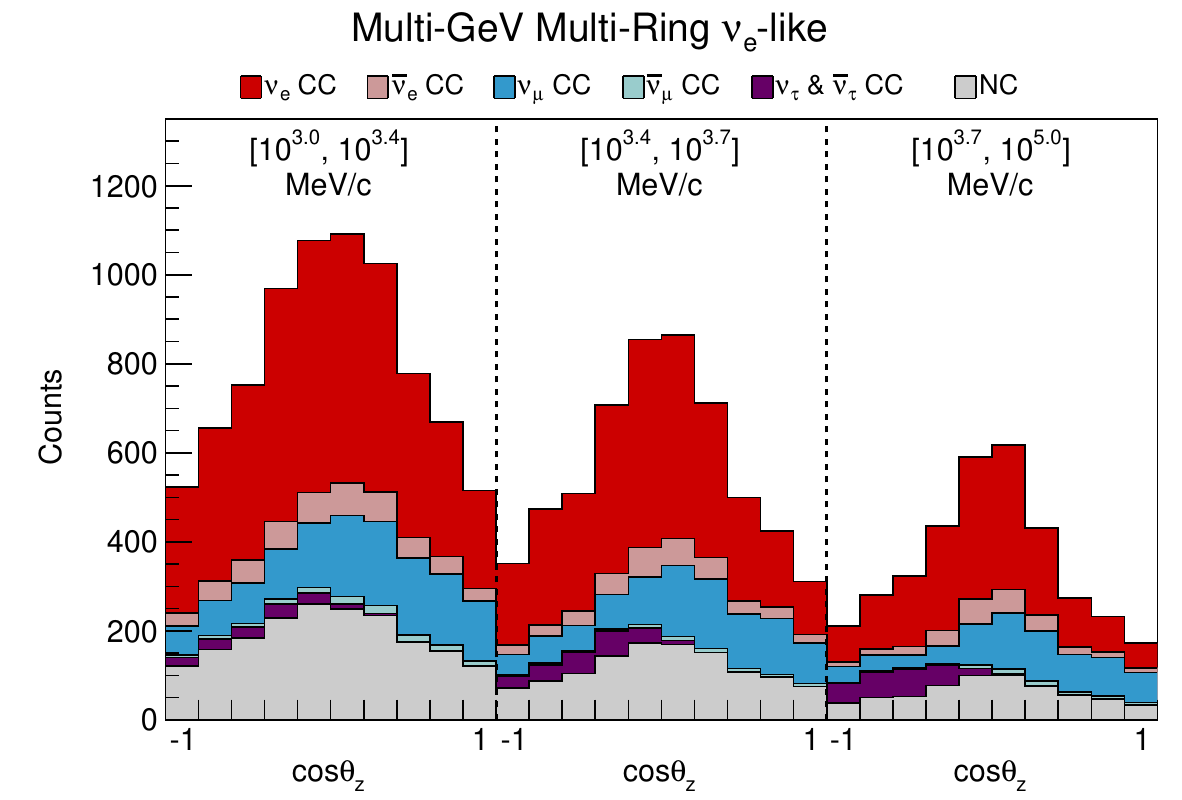}
\includegraphics[width=0.32\textwidth]{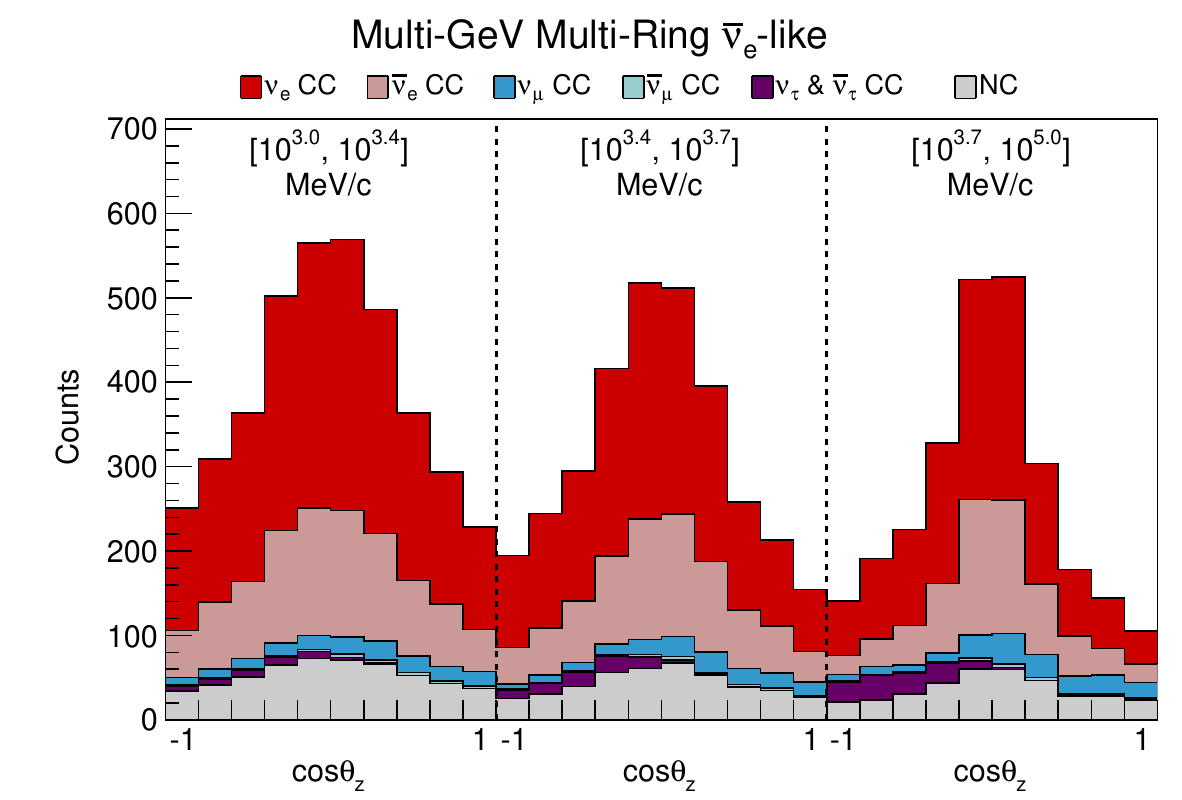}
\includegraphics[width=0.32\textwidth]{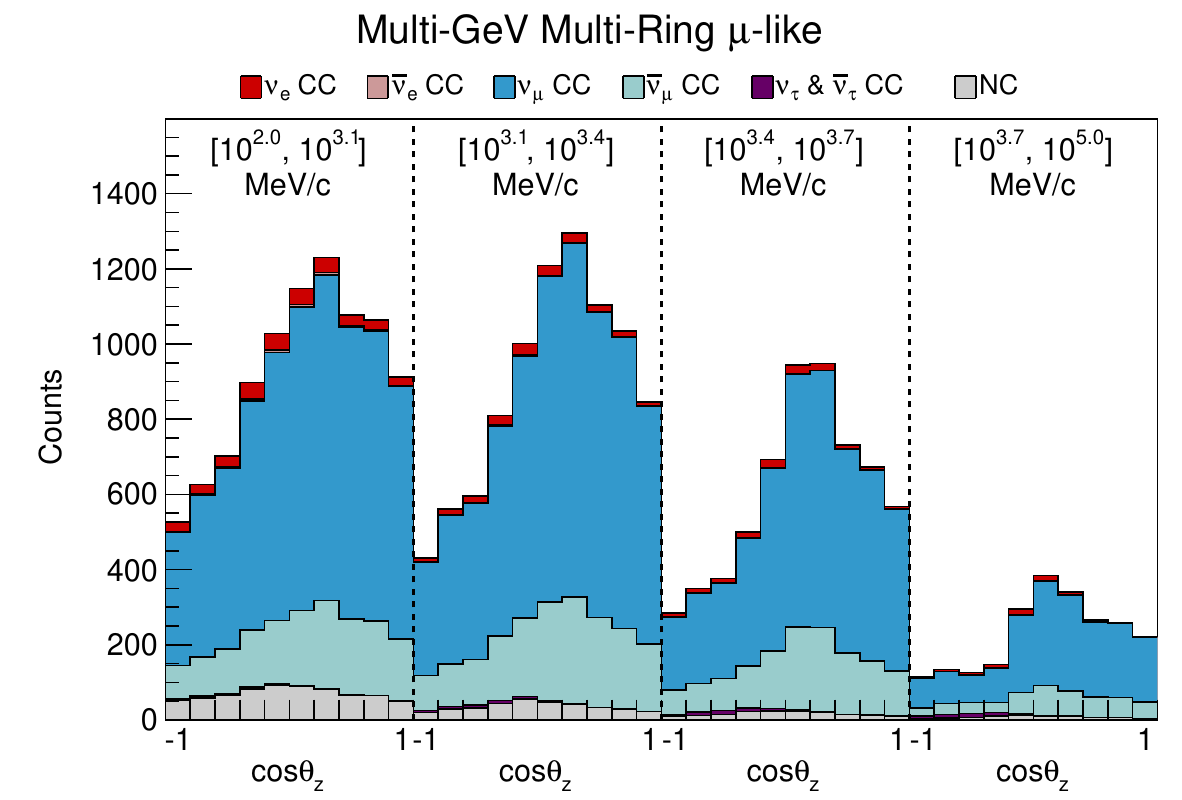}
\includegraphics[width=0.32\textwidth]{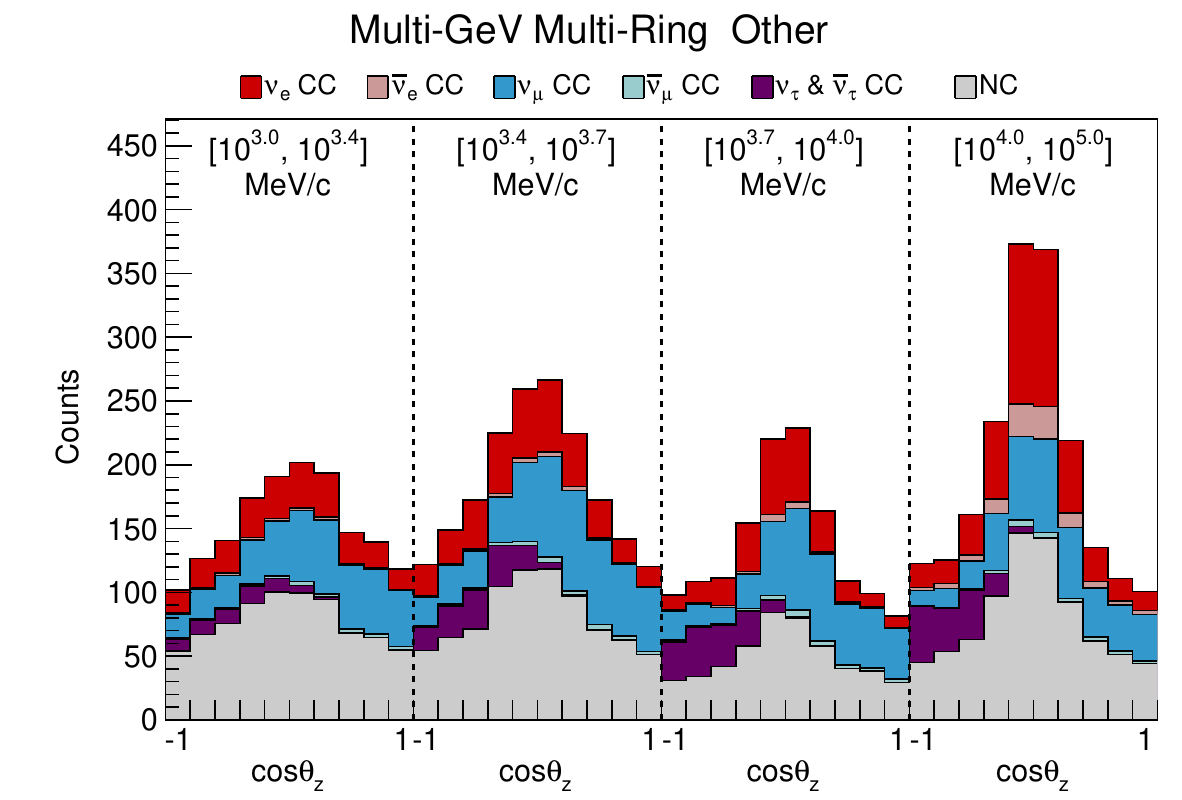}
\includegraphics[width=0.32\textwidth]{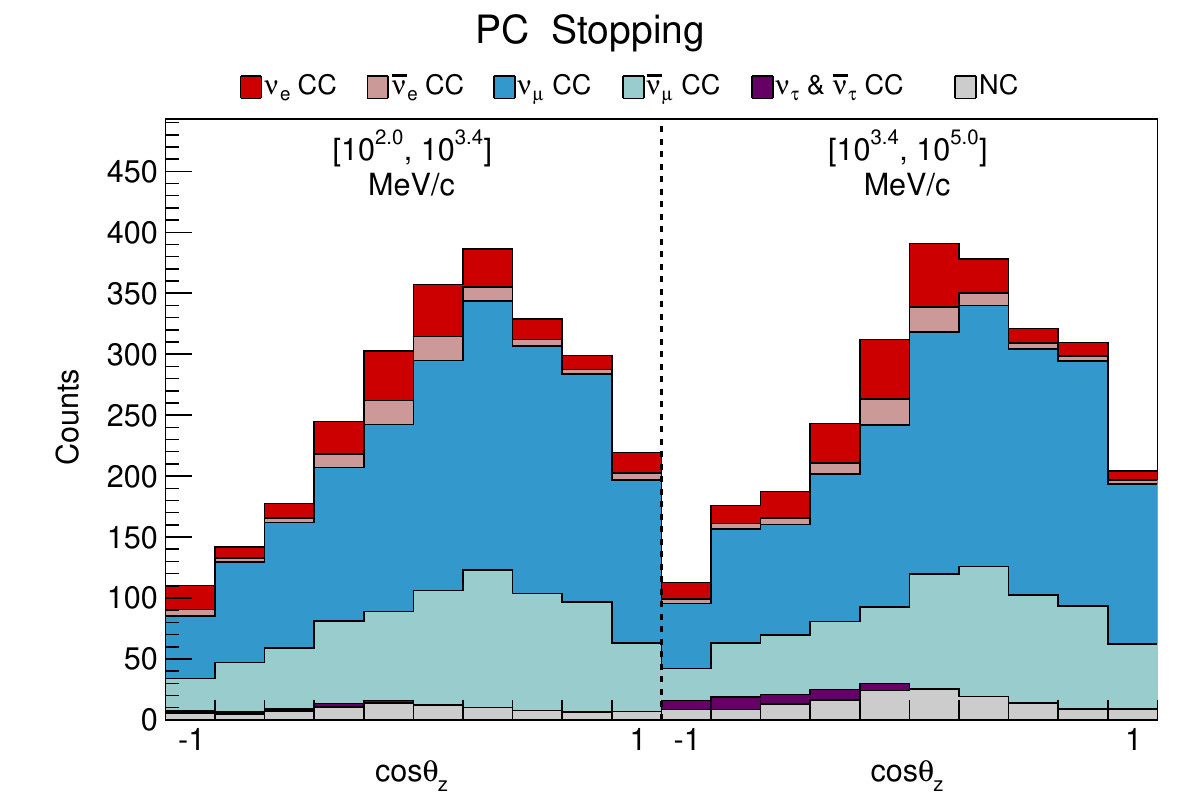}
\includegraphics[width=0.32\textwidth]{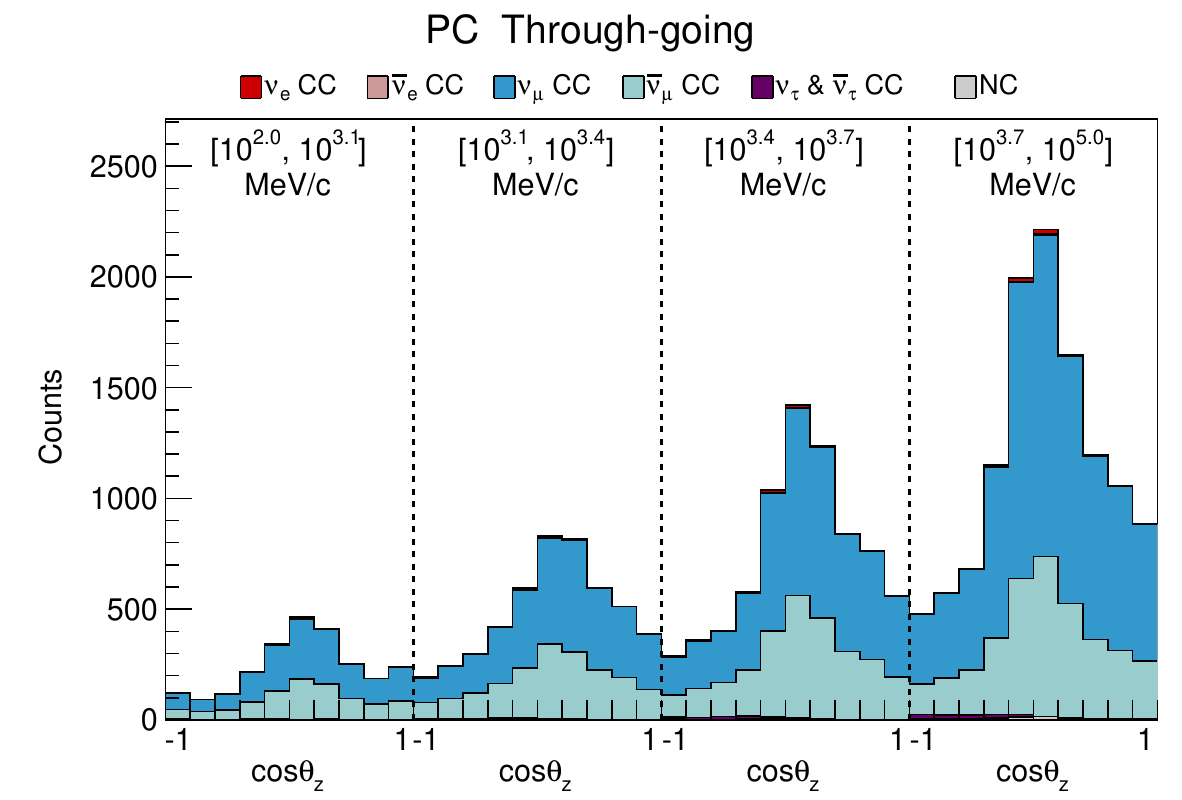}
\includegraphics[width=0.32\textwidth]{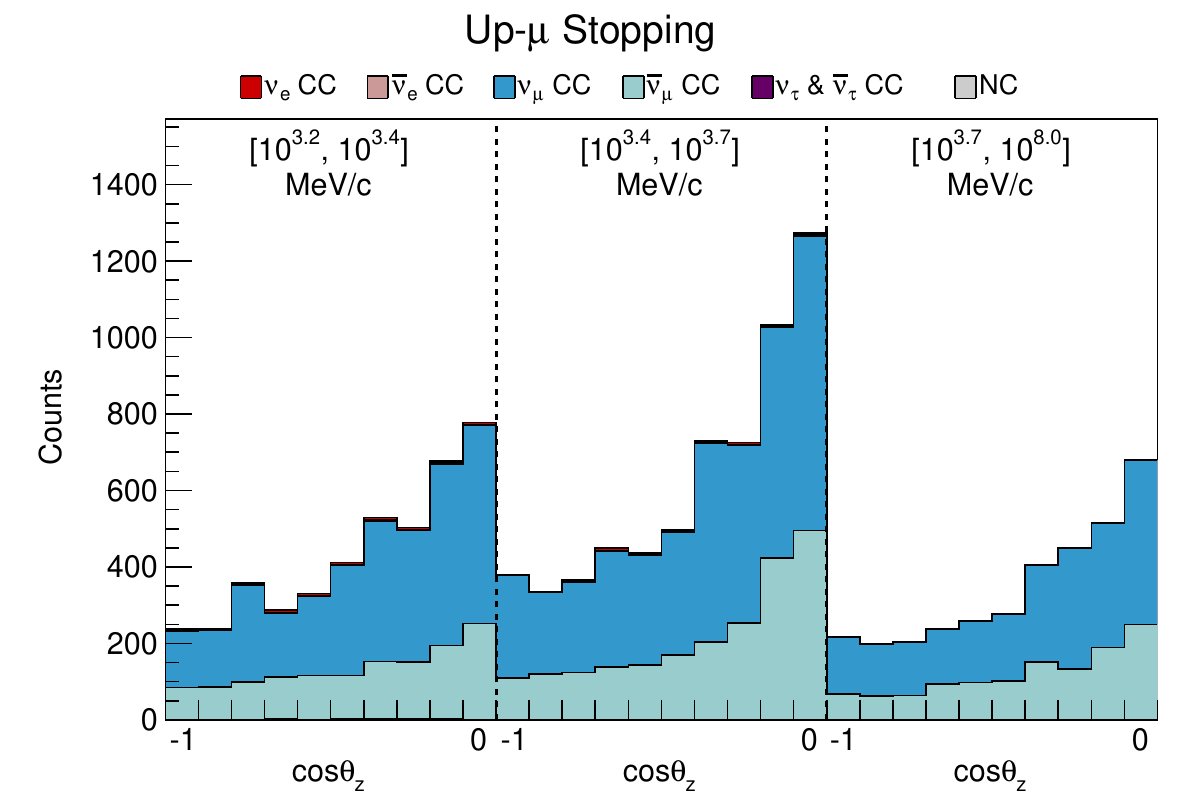}
\caption{Oscillated predictions, using parameters in Tab.~\ref{tab:sk_bfp}, 
for all analysis samples scaled to 6500-live days and considering a 
$\times$8 increased mass to scale Super-Kamiokande to Hyper-Kamiokande. 
All 2D distributions are presented in reconstructed lepton momentum slices.}
\label{fig:SelSamples}
\end{figure*}
The 29 selections samples can be sub-divided in 19 samples without and 10 
with neutron tagging, often indicated with additional labels $0$~n, $1$~n, etc. 
This large number of samples is the result, in part, to splitting some 
samples in two phases of the Super-Kamiokande experiment, i.e. data from SK 
runs 1 to 3 and from runs 4 to 5, further details on the specifics of 
this data periods are availabel in Ref.~\cite{Super-Kamiokande:2023ahc}. 
As we are primarily interested in Hyper-Kamiokande, whenever this duplicity 
exist we use the performances of the simulations for runs 4 and 5. 

Broadly speaking there are 4 different categories of samples: 
Sub-GeV samples, Multi- GeV samples, NC-samples, and non-fully-contained 
(non-FC) samples. Oscillations of neutrinos with energy below 1 GeV change very rapidly with energy and azimuth angle, see previous Fig. 2, such that tomography-sensitive variations get largely averaged out. Therefore
we do not consider Sub-GeV samples.
Similarly, as our interest regards effects in the neutrino flux 
flavor composition, we do not include NC-samples, insensitive to flavor 
transitions. Lastly, we only consider those samples that have a significant 
fraction of events with energies of interest. Accordingly, we ignore the 
Up-$\mu$ showering and Up-$\mu$ stopping samples, which largely consist 
of events with energies above several tens of GeV. 

Therefore, in this analysis we utilize a total of 12 selection samples, 
consisting of Multi-GeV events with a single lepton ring ($\nu_\mu$, 
$\bar{\nu}_\mu$, $\nu_e$, $\bar{\nu}_e$~0~n, $\bar{\nu}_e$~1~n) and 
Multi-GeV events with multiple rings ($\nu_\mu$, $\nu_e$, $\bar{\nu}_e$ and 
other), and non-FC events (partially contained stopping, partially contained 
through-going and Up-$\mu$ stopping). All the analysis samples under 
consideration are presented in Fig.~\ref{fig:SelSamples}.

%
\subsection{Neutrino Fluxes}
%

We use the same neutrino flux used by the Super-Kamiokande collaboration 
to generate its expected event counts, namely we utilize the flux prediction 
of Honda et. al.~\cite{Honda:2011nf}. In the unoscillated prediction, 
a tiny flux of $\nu_\tau+\bar{\nu}_\tau$ is expected. If one would run 
this nominal expectation through Super-Kamiokande's event production and 
analysis pipeline, only a very small statistical sample of 
$\nu_\tau+\bar{\nu}_\tau$ would be obtained, making it difficult to evaluate 
through simulations the reconstruction and selection performances and 
efficiencies in each sample for this flavor. To solve this in a simple manner, 
the Super-Kamiokande collaboration calculates the expected  
$\nu_\tau+\bar{\nu}_\tau$ event rate using for it Honda et. al. 
model's $\nu_\mu$ flux, and later down-scales the events to their appropriate 
value. We take this method into account in our calculations.

%
\subsection{Oscillation probability evaluation}
%

The three-flavour oscillation probability is evaluated using the 
\texttt{Prob3++} framework 
\footnote{For detaila see https://github.com/rogerwendell/Prob3plusplus.}, 
developed by Super-Kamiokande collaborators. 
The framework calculates the oscillation 
probability including matter effects for any neutrino flavor transition 
as a function of the input PMNS parameters, the values of the neutrino mass 
squared differences  values, and the Earth model configuration, which is 
input as a collection of density values along the Earth radius. 
\texttt{Prob3++}  takes as input the Earth density profile given by a 
collection of points. We evaluate the Earth density at 82 points along 
the Earth radius. A subset of 61 points are distributed from 
0 to 5600~km every 100~km, with few additional points near the shell 
transitions. The remaining points describe the 5600~km to 6371~km region, 
sampled more frequently at irregular intervals to reflect the finner structure 
of the Earth crust.

The oscillation probability is calculated in every bin $i$ and for every 
neutrino flavour $\alpha$. 
The  oscillated and nominal expected number 
of counts for the $\alpha$ flavor in the $i$-th bin, $N^{osc}_{i,\alpha}$ 
and  $N^{nominal}_{i,\alpha}$, are related as follows:
\begin{align}
    N^{osc}_{i,\alpha} = N^{nominal}_{i,\alpha} 
\times \Phi^{osc}_{i,\alpha}/\Phi^{nominal}_{i,\alpha},
\end{align}
%
where $\Phi^{osc}_{i,\alpha}$ and $\Phi^{nominal}_{i,\alpha}$ represent the 
oscillated and nominal fluxes of  $\nu_\alpha$ in the  $i$-th bin.
To calculate the nominal flux, we evaluate the Honda flux 
model for the $\alpha$ flavor at the average neutrino energy and 
average zenith angle for the charged-current events associated to that 
flavor in the bin.

\begin{figure}[htp!]
\centering
\includegraphics[width=0.23\textwidth]{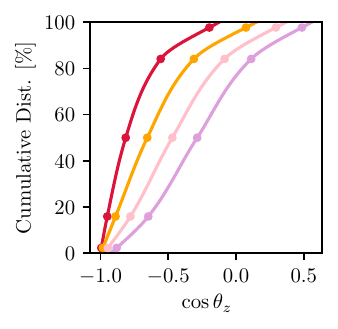}
\includegraphics[width=0.23\textwidth]{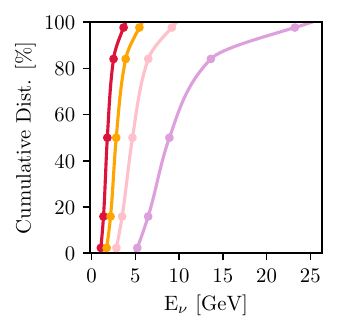}
\caption{Example splines in four consecutive energy bins (right panel) and 
four consecutive angle bins (left panel) in the 
{\rm Multi-GeV Multi-Ring} $\mu$-like 
sample for the $\nu_\mu$ reaction.
}
\label{fig:quant_examples}
\end{figure}

Event-by-event reweighting for neutrino oscillation effect is implemented in 
SK. However, the data and simulation releases consist of a simpler 
representation that loses the event-by-event details. Therefore, we instead 
conduct a quantile-weighted evaluation of the oscillation probability in every 
bin. Example spline distributions are presented in 
Fig.~\ref{fig:quant_examples}.
%

The procedure is as follows:
\begin{itemize}
    \item We use Piecewise Cubic Hermite Interpolating Polynomial 
(PHCIP)~\cite{fritsch1984method} monotonic splines to interpolate the quantiles.
    \item We define a number of equally spaced points, $N_q\in[0,1]$, 
forming a homogeneously spaced $N_q\times N_q$ grid. For every bin then 
we calculate the probability as a weighted sum:
  \begin{align}
    &\Phi^{osc}_{i,\alpha} = \Phi_\alpha \times \left(\sum^{N_q}_i \sum^{N_q}_j w_i
\times w_j \times P^{\alpha\rightarrow\alpha}_{osc}(E_i, \theta_j)\right)\nonumber\\
    &+\sum_{\beta \neq \alpha}\Phi_\beta \times \left(\sum^{N_q}_i \sum^{N_q}_j w_i
\times w_j \times P^{\beta\rightarrow\alpha}_{osc}(E_i, \theta_j)\right)\,,
\end{align}
%
where $w_i$, $w_j$ denote the width of the interval between two 
quantile points, with associated energy and azimuth angle values $E_i$ and 
$\theta_j$ evaluated at the center of that interval, 
 $P^{\beta\rightarrow\alpha}_{osc}(E_i, \theta_j)$ and
$P^{\alpha\rightarrow\alpha}_{osc}(E_i, \theta_j)$ being  
respectively the  $\nu_\beta \rightarrow \nu_\alpha$ transition 
and the $\nu_\alpha$ survival oscilation probabilities. 
\end{itemize}
For increasing $N_q$ the quantile intervals become smaller, closely 
reproducing the energy and angle distributions and ensuring a small variation 
of the oscillation probability from point to point. We use $N_q=15$ as we 
observed that finner grid sizes had no observable impact on the results. 

%
\subsection{Analysis datasets}
\label{sec:datasets}
%

We explore several scenarios for different assumptions of systematic 
uncertainties. In all cases we consider the same PID capabilities as in SK.

First, we consider the expected sensitivity for SK \texttt{SK-Nominal}, 
directly utilizing SK distributions and expected bin counts as in 
\cite{Super-Kamiokande:2023ahc} for the sub-set of samples considered in 
this work. Similarly, we report expected sensitivities scaled to 
HK's effective detection volume and 6500 days of data 
 (HK-Nominal), the  same as in Ref.~\cite{Super-Kamiokande:2023ahc}.

Second, we consider the expected sensitivity for sets of fake data informed by 
SK's performances. In total we consider the following detector performances:\\
\textbf{1)} E$_{0\%}$  \& 10$^\circ$ \textbf{2)} E$_{0\%}$ \& 20$^\circ$; \textbf{3)} E$_{30\%}$  \& 0$^\circ$ \textbf{4)} E$_{30\%}$ \& 10$^\circ$ \textbf{5)} E$_{20\%}$ \& 10$^\circ$ \textbf{6)} E$_{10\%}$ \& 10$^\circ$ \textbf{7)} E$_{30\%}$ \& 20$^\circ$.\\
The notation E$_{X\%}$ \& Y$^\circ$ indicates a $E_\nu$ resolution 
of X\% and a neutrino zenith angle resolution of Y$^\circ$.

Lastly we consider a special case:  E$_{0\%}$  \& 0$^\circ$, an upper bound 
for the sensitivity given only by statistical uncertainty. 
We refer to it as \textit{Stats.}.

%
\subsection{Enhanced Datasets Generation}
\label{fds_generation}
%

To generate the alternative datasets with enhanced detector performances, 
we sample individual events for every flavor ($\nu_e$, $\nu_\mu$, $\bar{\nu}_e$,
 $\bar{\nu}_\mu$, $\nu_\tau + \bar{\nu}_\tau$ and NC) from the PCHIP interpolated true energy and angle quantiles, thus producing un-binned 
distributions. In total, 10k events are sampled from every true energy 
and zenith angle bin of every sample. Each event weight corresponds to 
the number of counts in its original bin normalized by the number of 
event samples (10k). The reconstructed energy and azimuth angle for each event 
is then calculated applying a Gaussian smearing to its true simulated energy 
and azimuth angle according to the assumed detector performances, and events 
are organized in energy and angle bins accordingly. The resulting predictions 
are used to calculate 1D true neutrino energy and zenith angle quantile 
distributions in every reconstructed bin, and are used to generate a new 
'fake release' following the very same format as that in 
Ref.~\cite{SuperKamiokande2023}. This allows to run the same analysis 
pipeline both or SK/HK-Nominal  
distributions and for the enhanced datasets. 
To profit from the increase in statistics and the improved detector 
performances, the enhanced datasets utilize a different binning scheme. 
In particular, we double the number of angular bins at regular intervals 
from $\cos\theta=-1$ to $\cos\theta=0$, and more finely segment the energy 
bins from 1 to 10 GeV/c. We ignore all bins with $\cos\theta>0$, corresponding 
to neutrinos that arrive to the detector from the sky, i.e. without crossing 
the Earth. The total number of bins for the samples under study gets increased 
from 390 to 903.

%
\subsection{Sensitivity Calculation}
\label{chi2calc} 
%

The sensitivity to the parameters of interest ($\kappa_{C}$, 
$\kappa_{lman}$, $ \kappa_{uman}$) is calculated evaluating the statistical 
consistency of the PREM model expectations with the predictions for a modified 
Earth according to those parameters, utilizing a fixed choice of PMNS 
parameters, summarized in Table.~\ref{tab:sk_bfp}. Specifically, we calculate 
the $\Delta \chi^2 = \sum_i (O_i-E_i)^2/E_i)$, with $E_i$ and $O_i$ 
corresponding to the expected PREM model event counts in the bin $i-th$, and 
$O_i$ corresponding to the expected observed events counts for a modified 
Earth model in that same bin.
We report results for three values of $\sin^2\theta_{23} = 0.45$, 0.50 and 0.58,
which essentially span the $3\sigma$ allowed range of values of
$\sin^2\theta_{23}$ obtained in the latest global analysis of the
neutrino oscillation data \cite{Esteban:2024eli}.

%
\section{Results}
%

\begin{figure*}[htp!]
\centering
\includegraphics[width=0.32\textwidth]{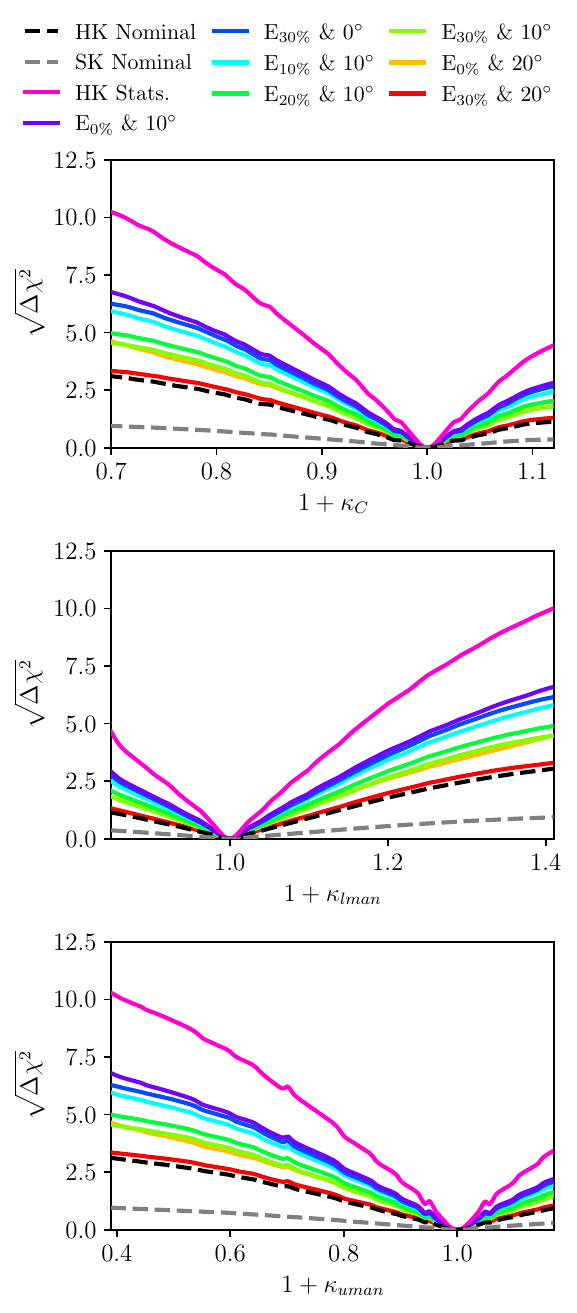}
\includegraphics[width=0.32\textwidth]{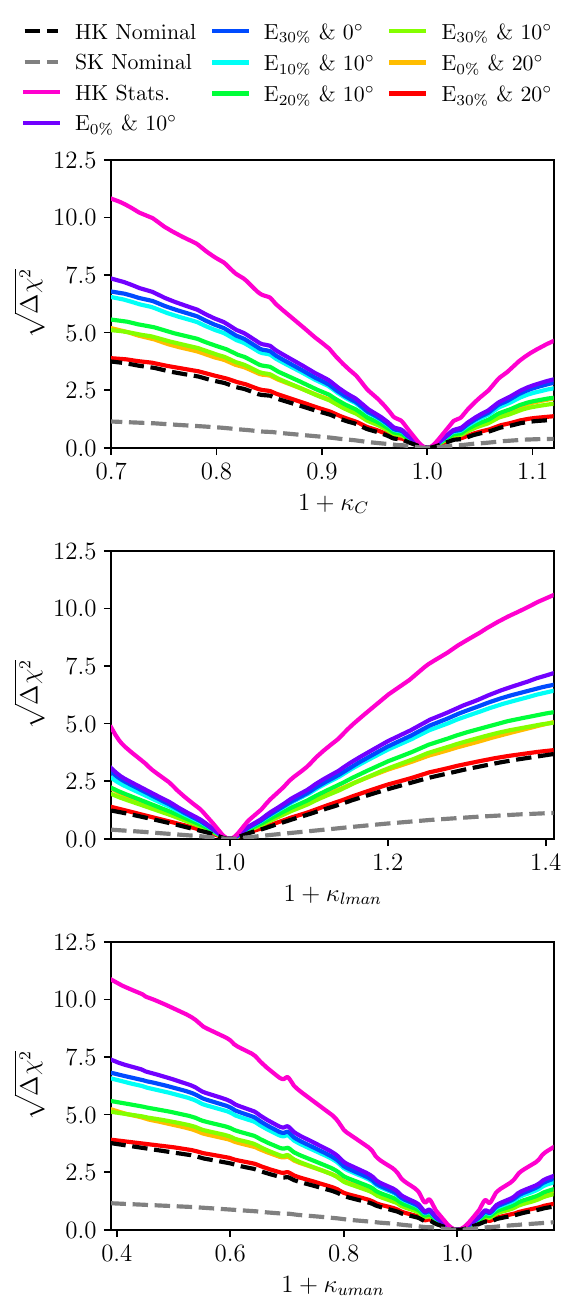}
\includegraphics[width=0.32\textwidth]{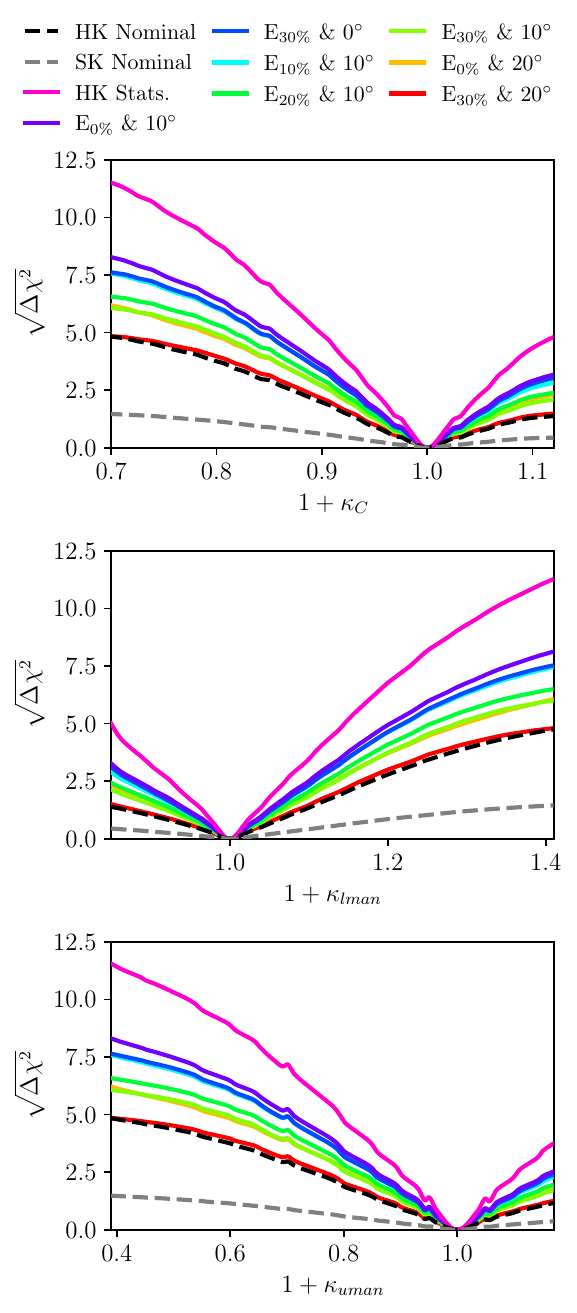}
\caption{Sensitivity to deviations   
of the Earth core, lower mantle and upper mantle densities
from their respective PREM values, parametrised
as $\rho_i(1 + \kappa_i)$, $i=c,lman,uman$,
$\rho_i$ being the PREM densities, for different HK detector configurations. 
The left, center and right columns correspond to: 
$\sin^2\theta_{23}=0.45$, 0.50 and 0.58. 
}
\label{fig:kappa_marginals}
\end{figure*}

\begin{figure}[htp!]
\centering
\includegraphics[width=0.45\textwidth]{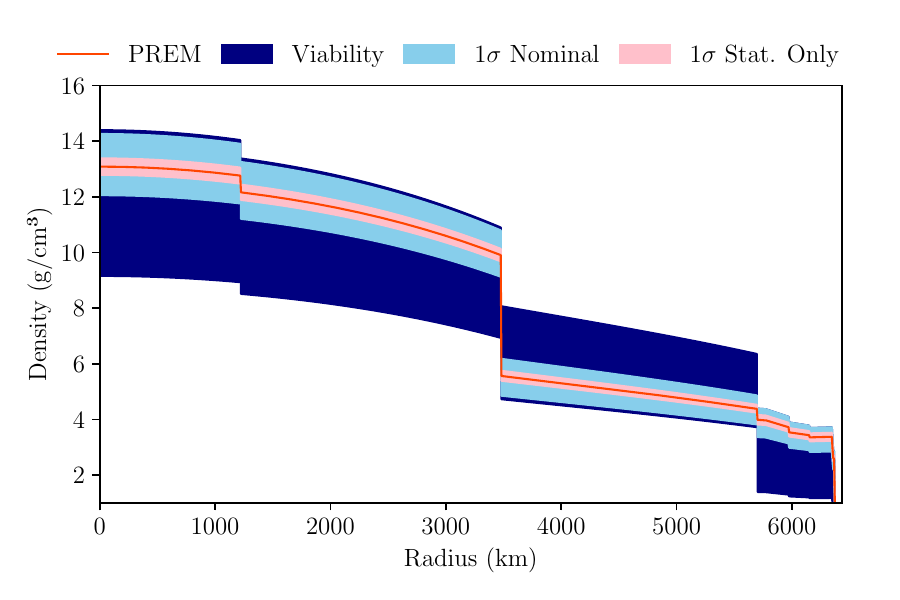}
\includegraphics[width=0.45\textwidth]{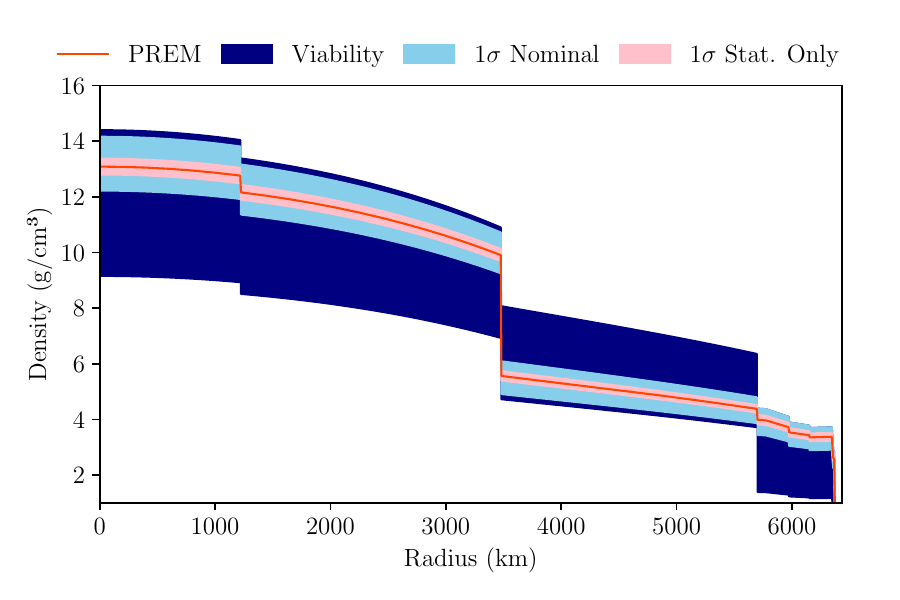}
\includegraphics[width=0.45\textwidth]{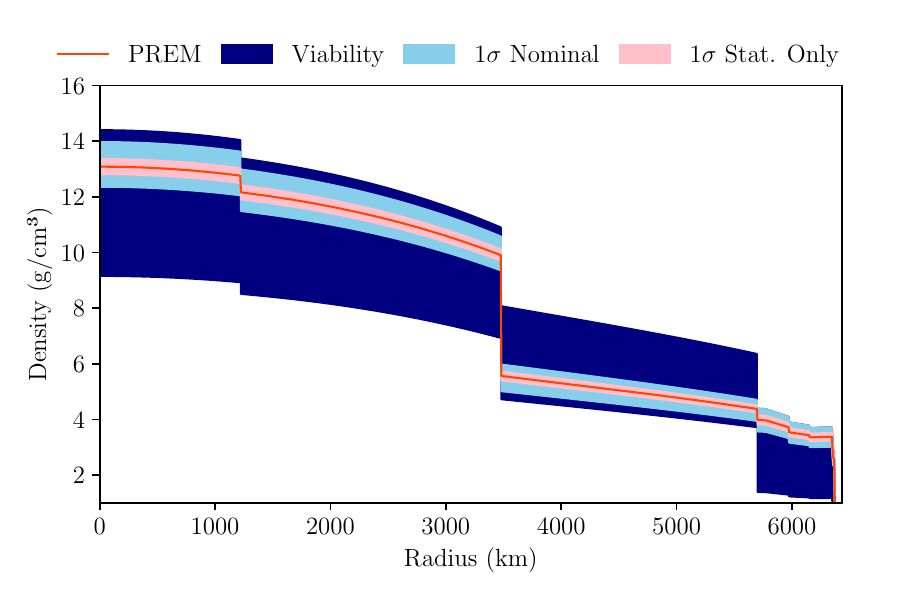}
\caption{Confidence intervals 
  of HK sensitivity to deviations of the
core, lower mantle and upper mantle average densities  
from their respective PREM values. The dark blue 
bands cosrrespond to varying $\kappa_C$ in the interval 
$0.30 < \kappa_C < 0.10$. The light blue and red bands 
represent the $1\sigma$ {\it Nominal} and {\it Stats.Only} 
confidence level intervals.
The top left, top right and bottom panels correspond respectively to 
$\sin^2\theta_{23}=0.45$, 0.50, 0.58.
See text for further details.
}
\label{fig:kappa_as_earth_profile}
\end{figure}

The sensitivity results for NO neutrino mass spectrum 
are presented in Fig.~\ref{fig:kappa_marginals}. 
Firstly, we note that the sensitivity of a detector with 
energy and zenith angle resolutions  
E$_{30\%}$ \& 20$^\circ$ in all samples is similar to 
the HK-Nominal sensitivity directly extrapolated from the performances 
of Super-Kamiokande. 
Correspondingly, we will refer to the case of 
E$_{30\%}$ \& 20$^\circ$ further in the text as ``nominal''.
From that starting point, we can notice that progressive improvements in the 
energy and angular resolutions would have a strong impact for a tomography of 
the Earth increasing the sensitivity to possible deviatons from the
PREM density values.
Our results suggest that enhancing the angular resolution
would translate more directly in tomography sensitivity enhancement.
For instance, the resolutions  
E$_{30\%}$ \& 10$^\circ$ and E$_{0\%}$ \& 20$^\circ$
provide very similar sensitivities,
while for E$_{30\%}$ \& 0$^\circ$ the sensitivity is much better.
In the case of having a 
better angular resolution, further improvement of the energy
reconstruction would of course have a positive impact,
as evidenced by the sensitivity in the 
E$_{20\%}$ \& 10$^\circ$ and E$_{10\%}$ \& 10$^\circ$ cases. Even in those 
scenarios, a notable margin to reach the ultimate sensitivity, given by the 
statistical only uncertainty case, exists - 
indicating that the measurement sensitivity is dominated by systematic uncertainties, 
as follows by comparing the 
results corresponding to E$_{0\%}$ \& 10$^\circ$
and HK {\it Stats}. An illustration of the extreme 
regimes of what is expected to be achieved and what could be best achieved 
in Hyper-Kamiokande is presented in Fig.~\ref{fig:kappa_as_earth_profile}. 
We show results for 
$\sin^2\theta_{23} = 0.45$, 0.50 and 0.58. The three values essentially 
span the $3\sigma$ allowed range of values of $\sin^2\theta_{23}$
obtained in the most recent global analysis of neutrino oscillation data 
\cite{Esteban:2024eli}. While comparing the results it is easy to 
notice that the true value of $\sin^2\theta_{23}$ has a significant 
impact on the HK sensitivity. 

 We summarise our results on the sensitivity of the 
HK experiment to possible deviations of the core,
lower mantle and upper mantle average 
densities from their respective PREM values 
in  Table~\ref{tab:comprehensive_summary}.

  We find, in particular, 
that for $\sin^2\theta_{23}$ 
and the ``nominal'' energy and zenith angle resolutions of 
E$_{30\%}$ \& 20$^\circ$ in all samples, after 6500 days of operation 
the HK experiment can determine the Earth core average density 
at $2\sigma$ C.L. with a relative uncertainty 
$\Delta \bar{\rho}_C/\bar{\rho}_C = 100\%\kappa_C$ 
of (-14.5\%)/+39.5 ($-0.145\leq \kappa_c\leq 0.395$).
For $\sin^2\theta_{23}= 0.58$ the uncertainty is noticeably smaller:
(-9.3\%)/+31.7\% ($-\,0.093\leq \kappa_c\leq 0.317$).
However, as we have shown, the $M_\oplus$, $I_\oplus$ and 
EHE constraints limit the possible maximal deviation of 
$\bar{\rho}_C$ from its PREM value to 10\% ($\kappa_c < 0.10$, 
see Eqs. (\ref{eq:ka3l}) and (\ref{eq:ka3lnum})).

The sensitivity of the HK experiment will be much higher 
in the case of energy and angular resolutions 
of E$_{20\%}$ \& 10$^\circ$. It increases also significantly 
with $\sin^2\theta_{23}$. Indeed, in this case at $2\sigma$ C.L. 
we have for $\sin^2\theta_{23} = 0.45$, 0.50 and 0.58, respectively:
(-9.2\%)/+11.3\% ($-\,0.092\leq \kappa_c\leq 0.113$),
(-8.3\%)/+9.8\% ($-\,0.083\leq \kappa_c\leq 0.098$) and 
(-6.7\%)/+8.5\% ($-\,0.067\leq \kappa_c\leq 0.085$).
Thus, if, e.g.,  $\sin^2\theta_{23} = 0.50$, the HK experiment 
would be able to establish that the average core density 
at $2\sigma$ C.L. lies in the rather narrow interval:
$10.08~{\rm g/cm^3} \leq \bar{\rho}_C 
\leq 12.06$ g/cm$^3$, with the PREM value of 
$\bar{\rho}_C = 10.99$  g/cm$^3$. 
 Similarly, it will follow from the HK data that 
at $2\sigma$ C.L. the average lower and upper mantle densities, 
have values in the intervals:
$4.49~{\rm g/cm^3} \leq  \bar{\rho}_{lman} \leq 5.38~{\rm g/cm^3}$, 
$3.00~{\rm g/cm^3} \leq  \bar{\rho}_{uman} \leq 4.25~{\rm g/cm^3}$,
where we have used the corresponding results in 
Table \ref{tab:comprehensive_summary} and 
the upper limit $\kappa_{uman} \leq 0.18$ (see Eq. (\ref{eq:ka3l})).   
We recall that  the PREM values of $\bar{\rho}_{lman}$ and $\bar{\rho}_{uman}$ 
are $4.90~{\rm g/cm^3}$ and  $3.60~{\rm g/cm^3}$, respectively.

 Given the PREM values of  $\bar{\rho}_{man} = 4.66$  g/cm$^3$ 
and $\bar{\rho}_{IC} = 12.89$  g/cm$^3$, 
the HK result on  $\bar{\rho}_{C}$
would be an independent confirmation of the exsitence of at least 
three major density layers in the interior of the Earth. 

\begin{table*}[htbp]
\centering
\resizebox{\textwidth}{!}
{
\begin{tabular}{|m{2.2cm}
|>{\centering\arraybackslash}m{1.2cm}
|>{\centering\arraybackslash}m{0.8cm}
|>{\centering\arraybackslash}m{0.8cm}
|>{\centering\arraybackslash}m{0.8cm}
|>{\centering\arraybackslash}m{0.8cm}
|>{\centering\arraybackslash}m{0.8cm}
|>{\centering\arraybackslash}m{0.8cm}
|>{\centering\arraybackslash}m{0.8cm}
|>{\centering\arraybackslash}m{0.8cm}
|>{\centering\arraybackslash}m{0.8cm}
|>{\centering\arraybackslash}m{0.8cm}
|>{\centering\arraybackslash}m{0.8cm}
|>{\centering\arraybackslash}m{0.8cm}
|}
\hline\hline
  & $\sin^2 \theta_{23}$ & \multicolumn{4}{c|}{Core (C)} &\multicolumn{4}{c|}{Lower Mantle} & \multicolumn{4}{c|}{Upper Mantle} \\
\hline
& & \multicolumn{2}{c|}{68\%} & \multicolumn{2}{c|}{95\%} 
  & \multicolumn{2}{c|}{68\%} & \multicolumn{2}{c|}{95\%} 
  & \multicolumn{2}{c|}{68\%} & \multicolumn{2}{c|}{95\%} \\
\hline
Model & & lower & upper & lower & upper & lower & upper & lower & upper 
      & lower & upper & lower & upper \\
\hline
\multirow{3}{*}{HK Nominal} 
& 0.45 & 0.922 & 1.090 & 0.834 & - & 0.873 & 1.111 & - & 1.228 & 0.844 & 1.184 & 0.671 & - \\
& 0.50 & 0.934 & 1.082 & 0.867 & - & 0.883 & 1.095 & - & 1.185 & 0.863 & 1.168 & 0.734 & - \\
& 0.58 & 0.945 & 1.067 & 0.899 & - & 0.902 & 1.074 & - & 1.145 & 0.895 & 1.141 & 0.793 & - \\
\hline
\multirow{3}{*}{SK Nominal}
& 0.45 & - & - & - & - & - & - & - & - & - & - & - & - \\
& 0.50 & 0.762 & - & - & - & - & 1.333 & - & - & 0.527 & - & - & - \\
& 0.58 & 0.829 & - & - & - & - & 1.237 & - & - & 0.658 & - & - & - \\
\hline
\multirow{3}{*}{HK Stats.}
& 0.45 & 0.978 & 1.023 & 0.953 & 1.048 & 0.969 & 1.033 & 0.934 & 1.064 & 0.958 & 1.043 & 0.906 & 1.096 \\
& 0.50 & 0.978 & 1.022 & 0.956 & 1.046 & 0.970 & 1.031 & 0.937 & 1.060 & 0.959 & 1.042 & 0.913 & 1.091 \\
& 0.58 & 0.979 & 1.021 & 0.959 & 1.043 & 0.972 & 1.029 & 0.940 & 1.056 & 0.960 & 1.041 & 0.922 & 1.084 \\
\hline
\multirow{3}{*}{E$_{0\%}$ \& 10$^\circ$}
& 0.45 & 0.963 & 1.038 & 0.930 & 1.075 & 0.950 & 1.049 & 0.894 & 1.102 & 0.932 & 1.069 & 0.854 & 1.150 \\
& 0.50 & 0.965 & 1.036 & 0.936 & 1.069 & 0.954 & 1.045 & 0.900 & 1.092 & 0.935 & 1.066 & 0.868 & 1.144 \\
& 0.58 & 0.969 & 1.033 & 0.942 & 1.064 & 0.960 & 1.041 & 0.910 & 1.078 & 0.939 & 1.063 & 0.890 & 1.129 \\
\hline
\multirow{3}{*}{E$_{30\%}$ \& 0$^\circ$}
& 0.45 & 0.961 & 1.040 & 0.924 & 1.079 & 0.945 & 1.053 & 0.888 & 1.108 & 0.927 & 1.075 & 0.848 & 1.159 \\
& 0.50 & 0.963 & 1.038 & 0.932 & 1.074 & 0.950 & 1.048 & 0.895 & 1.099 & 0.933 & 1.069 & 0.857 & 1.149 \\
& 0.58 & 0.966 & 1.035 & 0.939 & 1.066 & 0.956 & 1.043 & 0.904 & 1.084 & 0.937 & 1.065 & 0.880 & 1.139 \\
\hline
\multirow{3}{*}{E$_{10\%}$\& 10$^\circ$}
& 0.45 & 0.959 & 1.043 & 0.919 & 1.086 & 0.941 & 1.056 & 0.877 & 1.114 & 0.922 & 1.083 & 0.839 & 1.179 \\
& 0.50 & 0.963 & 1.040 & 0.930 & 1.080 & 0.946 & 1.050 & 0.886 & 1.102 & 0.931 & 1.073 & 0.854 & 1.162 \\
& 0.58 & 0.966 & 1.036 & 0.939 & 1.070 & 0.954 & 1.043 & 0.898 & 1.085 & 0.937 & 1.066 & 0.880 & 1.145 \\
\hline
\multirow{3}{*}{E$_{20\%}$\& 10$^\circ$}
& 0.45 & 0.952 & 1.051 & 0.908 & 1.113 & 0.930 & 1.066 & 0.855 & 1.135 & 0.904 & 1.101 & 0.803 & - \\
& 0.50 & 0.957 & 1.046 & 0.917 & 1.098 & 0.937 & 1.058 & 0.864 & 1.117 & 0.918 & 1.092 & 0.833 & - \\  
& 0.58 & 0.963 & 1.040 & 0.933 & 1.085 & 0.945 & 1.049 & 0.879 & 1.098 & 0.933 & 1.075 & 0.859 & 1.175 \\
\hline
\multirow{3}{*}{E$_{30\%}$ \& 10$^\circ$}
& 0.45 & 0.945 & 1.059 & 0.894 & - & 0.921 & 1.074 & - & 1.150 & 0.895 & 1.112 & 0.788 & - \\
& 0.50 & 0.953 & 1.054 & 0.909 & - & 0.928 & 1.065 & - & 1.131 & 0.905 & 1.104 & 0.808 & - \\
& 0.58 & 0.960 & 1.045 & 0.924 & 1.106 & 0.937 & 1.054 & 0.859 & 1.107 & 0.926 & 1.091 & 0.849 & - \\
\hline
\multirow{3}{*}{E$_{0\%}$\& 20$^\circ$}
& 0.45 & 0.949 & 1.052 & 0.896 & 1.116 & 0.929 & 1.070 & 0.854 & 1.148 & 0.899 & 1.102 & 0.790 & - \\
& 0.50 & 0.955 & 1.048 & 0.911 & 1.104 & 0.934 & 1.062 & 0.860 & 1.129 & 0.910 & 1.096 & 0.813 & - \\
& 0.58 & 0.961 & 1.042 & 0.926 & 1.090 & 0.942 & 1.052 & 0.871 & 1.105 & 0.929 & 1.081 & 0.851 & 1.185 \\
\hline
\multirow{3}{*}{E$_{30\%}$\& 20$^\circ$}
& 0.45 & 0.931 & 1.079 & 0.855 & - & 0.889 & 1.100 & - & 1.202 & 0.856 & 1.157 & 0.711 & - \\
& 0.50 & 0.939 & 1.072 & 0.880 & - & 0.897 & 1.085 & - & 1.169 & 0.879 & 1.147 & 0.760 & - \\
& 0.58 & 0.950 & 1.063 & 0.907 & - & 0.911 & 1.068 & - & 1.136 & 0.900 & 1.128 & 0.802 & - \\
\hline\hline
\end{tabular}
}
\caption{
The 68\% and 95\% C.L. intervals of sensitivity of the HK experiment 
to the parameters $(1 + \kappa_{C})$, $(1 + \kappa_{lman})$ 
and $(1 + \kappa_{uman})$,
characterising the deviations of the inner and outer core,
lower mantle and upper mantle densities from their respective PREM values,
for different sets of HK energy and angular resolutions and 
$\sin^2\theta_{23} = 0.45$, 0.50 and 0.58. 
Dashes indicate the limits obtained from the Earth $M_\oplus$, $I_\oplus$ 
and EHE constraints quoted in Eq. (\ref{eq:ka3lnum}) 
in the cases when they are more constraining than those
derived from the analysis of the HK prospective data.
See text for further details.
}
\label{tab:comprehensive_summary}
\end{table*}

%
\section{Conclusions}
%

  In the present study we have investigated the potential of the
Hyper-Kamikande experiment to perform neutrino tomography
of the Earth density structure. Using PREM as a reference model for
the  Earth density distribution we studied the sensitivity of
the Hyper-Kamiokande detector to deviations of the Earth 
total core average density $\bar{\rho}_C$, lower mantle
average density $\rho_{lman}$, and upper mantle average
density, $\bar{\rho}_{uman}$, from their respective PREM values.
The analysis is performed by studying the effects of the Earth matter 
on the oscillations of atmospheric $\nu_{\mu}$, $\nu_e$, $\bar{\nu}_\mu$ 
and $\bar{\nu}_e$. We used the public release~\cite{SuperKamiokande2023}
of the  simulated atmospheric neutrino event rate predictions generated by the 
Super-Kamiokande (SK) collaboration in its latest  atmospheric neutrino 
oscillation analysis~\cite{Super-Kamiokande:2023ahc},
which includes
i) realistic atmospheric neutrino flux
and neutrino interaction modeling,
ii) detector effects, and
iii) analysis of the sample selection criteria.
Based on those simulations, we extrapolated the Earth tomography sensitivity
to Hyper-Kamiokande (HK), a detector eight times larger than SK, but based on 
the same detection concept and with small instrumentation differences. 
Additionally, we modified the nominal distributions
(as discussed in Sec.~\ref{fds_generation})
to assess the potential impact of enhanced 
energy and/or angular resolutions in HK.
%

In investigating the potential of the HK experiment in doing a tomography 
of the Earth we have considered 6500 days of data, the same as in the latest 
study by Super-Kamiokande [72].  A significant gap is observed between the 
Nominal (or ``nominal'') sensitivity and the sensitivities achieved when 
using an improved angular and energy resolution 
(Fig.~\ref{fig:kappa_as_earth_profile}).
Therefore, reconstruction improvements from SK to HK, rooted for instance 
in machine learning techniques, can have a significant impact on the prospects 
of the HK experiment when performing a neutrino tomography of the Earth.

In our analysis we implemented the constraints on the variations of 
$\rho_C$, $\rho_{lman}$ and $\rho_{uman}$ 
following from the precise knowledge of the 
Earth mass $M_\oplus$ and moment of inertia $I_\oplus$, as well as from 
the requirement that the Earth be in hydrostatic equilibrium (EHE).

 Considering the case of normal ordering (NO) of neutrino masses,
we obtained results which illustrate the dependence of 
sensitivity to the core, lower and upper matle densities 
on the energy and zenith angle resolutions, on whether or not 
the prospective systematic errors are accounted for  
and on the value of the  atmospheric neutrino mixing 
angle $\theta_{23}$ (Fig.~\ref{fig:kappa_marginals},
Table \ref{tab:comprehensive_summary}).
We have shown, in particular, that a progressive improvement of the
``nominal'' energy and angular resolutions of
E$_{30\%}$ \& 20$^\circ$ of the HK detector
would have a strong impact on the Earth tomography potential
of the HK experiment. Our results suggest that
enhancing the angular resolution has a particularly strong impact
increasing the sensitivity of the HK experiment to possible deviations 
of the considered Earth layers average 
densities from their respective PREM values. 

 Performing the amalysis for three values of
$\sin^2\theta_{23} = 0.45$, 0.50 and 0.58,
which essentially span the $3\sigma$ allowed range of values of
$\sin^2\theta_{23}$ obtained in the latest global analysis of the
neutrino oscillation data, we have shown also that
the Earth tomography potential of the HK experiment 
depends strongly on the value of $\sin^2\theta_{23}$.

 More specifically, 
in the case of the ``nominal'' 
$E_{res} = 30\%$, $\theta_{zres} = 20^\circ$
and, e.g., $\sin^2\theta_{23}=0.45~(0.58)$, the 
Hyper-Kamiokande can determine the average core  
density $\bar{\rho}_C$ at $2\sigma$ C.L. after 6500
days of operation with an the uncertainty of (-14.5\%)/+39.5\% ((-9.3\%/+31/7\%),
while the $M_\oplus$, $I_\oplus$ and EHE constraints 
allow in the case of the three layer Earth density structure considered by us 
a maximal increase of $\bar{\rho}_C$ 
with respect to its PREM value of less than $\sim 10\%$.
In the ``more favorable'' case of implemented constraints, 
neutrino energy resolution $E_{res}$  of 20\%,  
zenith angle resolution $\theta_{zres}$ of 10\% 
and $\sin^2\theta_{23}=0.58~(0.45)$,
the HK experiment can determine the average core  
density $\bar{\rho}_C$ at $2\sigma$ C.L.  
with an uncertainty of (-8.3\%)/+9.8\%  (of (-9.2\%)/+11.3\%).  
If, e.g.,  $\sin^2\theta_{23} = 0.50$, the HK experiment 
would be able to establish, in particular, that the average core density 
at $2\sigma$ C.L. lies in the rather narrow interval:
$10.08~{\rm g/cm^3} \leq \bar{\rho}_C 
\leq 12.06$ g/cm$^3$, with the PREM value of 
$\bar{\rho}_C = 10.99$  g/cm$^3$. 
Given the PREM values of  the average densities of
the mantle and the inner core, $\bar{\rho}_{man} = 4.66$  g/cm$^3$ 
and $\bar{\rho}_{IC} = 12.89$  g/cm$^3$, 
this result would  be an independent 
evidence of the existence of at least 
three major density layers in the interior of the Earth and 
of the relatively large jump of density in the mantle-core narrow 
transition zone. 

Our results show that the Hyper-Kamiokande experiment,
after collecting a sufficiently large data sample, 
can make a major contribution  
to the tomography studies of the
Earth interior with neutrinos.

\section*{Acknowledgements}

We would like to note that the present study is not a study performed by the
Hyper-Kamioknade collaboration although in it we discuss one specific physics
problem ---that of getting information about the Earth density structure--- which
can be addressed by the Hyper-Kamioknade experiment.
The authors are indebted to Thomas Wester for clarifications on 
the assumptions and formatting of the Super-Kamiokande's data release.
C.J.V. was supported in part by JSPS KAKENHI Grant-in-Aid JP24K17065. 
The work of S.T.P. was supported in part by the European 
Union's Horizon 2020 research and innovation programme under 
the Marie 
Sklodowska-Curie grant agreement No.~860881-HIDDeN, 
by the INFN program on Theoretical Astroparticle Physics 
and by the  World Premier International Research Center
Initiative (WPI Initiative, MEXT), Japan.
S.T.P would like to thank Kavli IPMU, University of Tokyo, where 
a major part of this study was done, for the kind hospitality.
J.X was supported by Grant-in-Aid for JSPS Fellows P22333.

\end{document}